\documentclass[a4paper,11pt]{article}
\usepackage{amsmath,amssymb,bm}
\usepackage{titlesec}
\titleformat{\paragraph}[runin]{\normalfont\itshape}{\theparagraph.}{.3em}{}[.]\titlespacing{\paragraph}{0pt}{1ex plus .1ex minus .2ex}{.5em}
\usepackage[letterpaper, hmargin=1in, top=1in, bottom=1.2in, footskip=0.6in]{geometry}
\usepackage[utf8]{inputenc}
\usepackage{lmodern}
\usepackage{mathtools}
\usepackage{dsfont}
\pdfoutput=1
\usepackage[T1]{fontenc}
\usepackage[english]{babel}
\usepackage{graphicx} 
\usepackage{booktabs} 
\usepackage{color}
\definecolor{aquamarine}{rgb}{0.5, 1.0, 0.83}
\definecolor{ao(english)}{rgb}{0.0, 0.5, 0.0}
\definecolor{armygreen}{rgb}{0.29, 0.33, 0.13}
\definecolor{awesome}{rgb}{1.0, 0.13, 0.32}
\definecolor{ballblue}{rgb}{0.13, 0.67, 0.8}
\definecolor{bittersweet}{rgb}{1.0, 0.44, 0.37}
\definecolor{blue}{rgb}{0.0, 0.0, 1.0}
\definecolor{brinkpink}{rgb}{0.98, 0.38, 0.5}
\definecolor{ballblue}{rgb}{0.13, 0.67, 0.8}
\definecolor{brightturquoise}{rgb}{0.03, 0.91, 0.87}
\definecolor{blue-green}{rgb}{0.0, 0.87, 0.87}
\definecolor{caribbeangreen}{rgb}{0.0, 0.8, 0.6}
\definecolor{cyan}{rgb}{0.0, 1.0, 1.0}
\definecolor{amber(sae/ece)}{rgb}{1.0, 0.49, 0.0}
\usepackage[utf8]{inputenc}
\usepackage{lmodern}
\graphicspath{ {images/} }

\author{J\"{u}rg Fr\"{o}hlich\footnote{Institut f\"ur Theoretiche Physik, ETH-Z\"urich , Switzerland / Email: juerg@phys.ethz.ch}\,\,\, and Alessandro Pizzo \footnote{Dipartimento di Matematica, Universit\`a di Roma ``Tor Vergata", Italy
/ Email: pizzo@mat.uniroma2.it}}

\title{The Time-Evolution of States in Quantum Mechanics}

\begin{document}

\maketitle

\begin{abstract}
It is argued that the Schr\"odinger equation does not yield a correct description of the quantum-mechanical time evolution of states of isolated (open) systems featuring events. A precise general law for the time evolution of states replacing the Schr\"odinger equation is formulated within the so-called $ETH$-Approach to Quantum Mechanics. This law eliminates the infamous ``measurement problem.'' Our general results are illustrated by an analysis of simple models describing a very heavy atom coupled to the quantized radiation field in a limit where the speed of light tends to infinity. The discussion of these models is the main subject of our paper.
\end{abstract}

\tableofcontents

\section{Introduction: In search of a new law of Nature}\label{Intro}

\hspace{0.5cm} \textit{``... their attempts to see in the very inadequacy of the conventional interpretation of quantum theory a 
deep physical principle have often led physicists to adopt obscurantist, mystical, positivist, psychical, and other irrational 
worldviews.''} (David Deutsch, \cite{Deutsch})\\

In this paper we attempt to add an important \textit{missing piece} to the puzzle of Quantum Mechanics (henceforth 
abbreviated as QM), namely an appropriate notion of states\footnote{with a clear ontological meaning} and a general {\bf{statistical law}} governing the \textit{time evolution of states} of isolated physical systems featuring events. Along the way we intend to dispose of the misconception 
that unitary Schr\"odinger evolution of unit rays in Hilbert space (or of density matrices) provides that missing piece.

Disagreement concerning the right notion of states in QM and the nature of a general law describing their time evolution 
has persisted for almost a century, despite various proposals of how to resolve it; see, e.g., \cite{Bohm, GRW, Everett}, 
and \cite{Bell, ABDZ} and references given there. This has perpetuated a never ending debate about the deeper meaning 
of QM and has caused a lot of confusion -- as deplored by \textit{David Deutsch}. 
Indeed, \textit{Sean Carroll} has expressed the following pessimistic assessment of the present level 
of understanding of Quantum Mechanics: \textit{``What we don't do is claim to $\mathbf{understand}$ 
quantum mechanics. Physicists don't understand their own theory any better than a typical smartphone 
user understands what's going on inside the device.''} (Sean Carroll, in: New York Times 2019)   

But the problem is \textit{not} that we may not have understood the deeper meaning of QM -- in other words that we may not 
have found the correct interpretation of QM, yet. The problem is that we have \textit{not} accomplished a \textit{complete 
formulation of the theory called Quantum Mechanics}, yet, as remarked, e.g., by \textit{Paul Adrien Maurice  Dirac} (see 
below)!  Perhaps, this is also what \textit{Richard Feynman} may have vaguely had in mind when he said: 
\textit{``I cannot define the real problem; therefore I suspect there's no real problem; but I'm not sure there's no real problem.''} 
-- The present state of affairs in our comprehension of Quantum Mechanics is undoubtedly most unsatisfactory, 
indeed, and should be changed for the better, as soon as possible! 

The main purpose of this paper is to not only formulate a general law describing the evolution of states in non-relativistic QM, 
but to exemplify it by analyzing a class of simple models of a very heavy atom coupled to the radiation field (in a limit where 
the speed of light tends to $\infty$), building on ideas described in \cite{F-Schub, BFS, Fr1, LMU-19}.  Although a superior theory, 
as compared to non-relativistic QM, \textit{local relativistic quantum theory} is  technically more complicated; and we do  
not  know any four-dimensional models of this theory with non-trivial interactions that have been shown to be mathematically 
consistent. The relativistic theory has been considered in \cite{Fr2} and will be studied in more detail in forthcoming work.

Here is a metaphor for the present situation and the task to be accomplished: Until now, QM rests on  only 
three pillars, to be recalled in the next section, and remains incomplete. Our task is to construct a \textit{Fourth Pillar}\footnote{During the Song Dynasty, X\'u Zi Píng reformed Li Xu-Zhong's \textit{``Three Pillars of Destiny''} by adding the \textit{birth time} as the \textit{``fourth pillar.''...} Source: wikipedia.org/wiki/Four Pillars of Destiny} that will 
make the foundations of QM solid and stable. Built into our approach towards accomplishing this task is the fundamental 
dichotomy of \textit{future}, as the realm of potentialities, versus \textit{past}, as the realm of actualities 
and facts. It thus incorporates ancient philosophical ideas, in particular \textit{Aristotle'}s distinction 
between potentialities and actualities.\\

Our paper is organized as follows.

In Sect.~\ref{three pillars}, some essential but standard elements of Quantum Mechanics are summarized. 
As usual, we represent physical quantities characteristic of a system by self-adjoint operators acting on a separable Hilbert 
space. We then emphasize the importance of the requirement that physical quantities can be localized in bounded intervals 
of the time axis. The Heisenberg picture and the usual Heisenberg equations for the time evolution of operators representing 
physical quantities characteristic of \textit{isolated} (open) physical systems are described. The Copenhagen Interpretation 
of Quantum Mechanics (including the ``collapse postulate'') is briefly recapitulated.

In Sect.~\ref{Schroedinger evol}, we review arguments, including a gedanken experiment discussed in more detail in 
\cite{FFS} (see also \cite{Schill}), which show that linear unitary Schr\"odinger evolution does \textit{not} describe the 
time evolution of states of isolated physical systems featuring events, such as systems performing measurements. Some historical remarks introduce our reasoning process.

Sect.~\ref{ETH} is devoted to a short summary of the so-called $ETH$-Approach to Quantum Mechanics,\footnote{where 
``$ETH$'' stands for ``\textit{Events, Trees} and \textit{Histories}.''}  which has 
been developed with the intention to provide a completion of Quantum Mechanics that gets rid of conundrums such 
as the so-called ``measurement problem.'' Further details concerning the $ETH$-Approach can be found in 
\cite{F-Schub, BFS, Fr1, LMU-19, Fr2}. Precise notions of \textit{potential events/potentialities} and of 
\textit{actual events/actualities} are introduced; (for earlier work concerning various notions of ``events'' in QM, see 
\cite{Rudolf, Blanchard}, and references given there). Our notions of \textit{potentialities} and \textit{actualities} reflect the 
fundamental dichotomy of \textit{future} and \textit{past}. A physically meaningful concept of states of isolated physical 
systems is proposed. We then describe the so-called \textit{Principle of Diminishing Potentialities} and offer a concise 
formulation of the \textit{Collapse Postulate}. These ingredients enable us to formulate a general \textit{Law} describing 
the time evolution of states of isolated physical systems.

It turns out that the Principle of Diminishing Potentialities can be understood to be a consequence of Huygens' Principle in 
local relativistic quantum theories with massless particles, which has been analyzed and used in important work by 
\textit{Detlev Buchholz} \cite{Buchholz}; see also \cite{BRob}. In Sect.~\ref{HP}, that connection is recalled. We then propose 
a simple semi-relativistic model (with discrete time) illustrating the $ETH$-Approach to Quantum Mechanics and, 
in particular, the Principle of Diminishing Potentialities. We also comment on the form this principle takes in the limit 
where the speed of light tends to infinity.

The most important section of this paper is Sect.~\ref{Model}. It is devoted to a rather detailed study of models illustrating 
the $ETH$-Approach. These models can be interpreted as describing a very heavy atom with a finite-dimensional 
Hilbert space of internal states coupled to a caricature of the quantized electromagnetic field, called $R$-field, 
arising in the limit of the speed of light tending to infinity. The models are chosen so as to minimize technical 
complexity, but not to loose essential aspects of the $ETH$-Approach. Time is chosen to be discrete, 
and operators representing physical quantities localized in bounded intervals of the time axis generate 
\textit{finite-dimensional} matrix algebras. For these models, an explicit law for the evolution of states is 
derived. We then discuss the main implications of this law in two limiting regimes: a regime where the atom is only 
very weakly coupled to the $R$-field and, as a consequence, linear unitary Schr\"odinger evolution is a good approximation  
to the true evolution of states; and a regime where the degrees of freedom of the atom are very strongly coupled to the 
degrees of freedom of the $R$-field and the evolution of states is well approximated by a classical Markov chain. 
The section concludes with an explanation of how, in these models, ``measurements'' can be described in a very 
natural way.

In Sect.~\ref{Conclusions}, we briefly comment on the \textit{ontology} that underlies a quantum-mechanical description 
of Nature according to the $ETH$-Approach. We then sketch how the models discussed in Sect.~\ref{Model} can be 
extended to non-relativistic models with a \textit{continuous} time. We observe that, in such models, the spectrum 
of the Hamiltonian is unbounded from above and below. Finally, we comment on the problem of understanding 
whether there are alternatives to Huygens' Principle in deriving the Principle of Diminishing Potentialities. 
A tantalizing conclusion of our analysis is that a quantum theory satisfying this principle, as well as the 
\textit{spectrum condition}, which says that the energy spectrum of the Hamiltonian of the theory must be 
bounded from below (i.e., contained in a half-bounded interval of the real line), appears to be necessarily a 
\textit{local relativistic quantum theory}.

\textit{Remark}: In this paper we do not review the quantum theory of indirect (weak) measurements, which is well developed, 
taking certain results in a theory of direct (projective) measurements and events for granted.
See \cite{M-K}, and \cite{BCJP} for recent  results and plenty of references.

{\bf{Acknowledgements}}. One of us (J.F.) thanks colleagues at Ludwig-Maximilian University (LMU) in Munich, and 
in particular \textit{Heinz Siedentop}, for having invited him to teach a crash course of roughly twenty-four hours on 
basic Quantum  Mechanics. This course, which took place in November/December of 2019, was a great opportunity 
to experiment with some of the material presented in our paper, in particular with the models discussed in Sect.~\ref{Model}. 
He also thanks several colleagues in Munich, including \textit{Detlef D\"urr, Erhard Seiler} and \textit{Heinz Siedentop}, 
for very useful discussions and the pleasure of enjoying their company. Numerous discussions with former collaborators, especially with \textit{Baptiste Schubnel}, and encouragement from \textit{Shelly Goldstein} have been important. We are grateful to \textit{Claudio Paganini} for some useful comments.

\section{Three of the four pillars Quantum Mechanics rests upon}\label{three pillars}

\hspace{0.5cm} \textit{``It seems clear that the present quantum mechanics is not in its final form.''} (Paul Adrien Maurice Dirac)\\

In this section we summarize a few well known basic facts about \textit{non-relativistic Quantum Mechanics}, focussing on the quantum-mechanical description of physical quantities characteristic of a physical system and their 
dynamics in the \textit{Heisenberg picture}. 

Since we consider non-relativistic quantum mechanics, with gravity turned off (or treated as an instantaneous interaction 
between particles, as conceived by Newton), we may assume that the concept of an \textit{absolute time}, 
$t\in \mathbb{R}$, parametrising evolution is meaningful; (see \cite{Fr2} for a sketch of a \textit{space-time} 
approach to local relativistic quantum theory). 

\subsection{The usual three pillars}
\hspace{0.5cm}\textit{``If you are receptive and humble, mathematics will lead you by the hand.''} (Paul Adrien Maurice Dirac)

In this subsection we recall some well known elements (or ``pillars'') of a quantum-mechanical description of Nature.\\

\textit{\underline{Pillar 1}: Physical quantities characteristic of a system.}

In quantum mechanics, a physical system, $S$, is characterized by a list of abstract self-adjoint operators,
\begin{equation}\label{phys quantities}
\mathcal{O}_{S}= \big\{\hat{X}_i = \hat{X}^{*}_i\,\vert\, i \in \mathfrak{I}_S\big\}\,,
\end{equation}
with $\mathfrak{I}_S$ a set of indices depending on $S$, where every operator $\hat{X} \in \mathcal{O}_S$ represents a 
physical quantity characteristic of $S$, such as the total momentum, energy or spin of all particles localized in some bounded 
region of physical space and belonging to an ensemble of (possibly infinitely many) particles constituting the system $S$.\footnote{In order to comprehend the notion of physical quantities underlying the analysis presented in this paper, the reader 
may find it useful to recall the description of, for example, a quantum gas, such as the electon gas in a metal or a gas of 
bosonic atoms, in the \textit{formalism of second quantization}. See also Sect. \ref{Model}.} The set $\mathcal{O}_S$ of selfadjoint operators does not have any interesting structure. It is usually not a (real) linear space, let alone an algebra. 

At every time $t$, there is a representation of 
$\mathcal{O}_S$ by selfadjoint operators acting on a separable Hilbert space $\mathcal{H}_S$:
\begin{equation}\label{Rep}
\mathcal{O}_S \ni \hat{X} \mapsto X(t)= X(t)^{*} \in B(\mathcal{H}_S)\,,
\end{equation}
where $B(\mathcal{H}_S)$ is the algebra of all bounded operators acting on $\mathcal{H}_S$. Usually, 
a physical quantity $\hat{X}\in \mathcal{O}_S$ can be localized in space and in time (\textit{Haag} \cite{Haag} speaks of ``local observables,'' \textit{Bell} \cite{Bell} of ``local beables''). It can be constructed by testing some hermitian operator-valued 
density, $\hat{\mathfrak{x}}(x,\tau)$, on space-time, such as a mass-, momentum-, energy- or spin density of a quantum gas, 
with a real-valued test function $h(\mathbf{x},\tau)$, yielding a self-adjoint operator:
\begin{equation}\label{phys quantity}
\hat{X}=F\Big[\int d^{3}x\, d\tau\, h(\mathbf{x},\tau)\, \hat{\mathfrak{x}}(\mathbf{x},\tau)\,\Big]\, \mapsto \,\,
X(t) := F\Big[ \int d^{3}x \, d\tau \,h(\mathbf{x},\tau)\, \mathfrak{x}(\mathbf{x}, \tau+ t)\,\Big]\,,
\end{equation}
for arbitrary $t\in \mathbb{R}$, where $F$ is some bounded continuous function on $\mathbb{R}$, and $\mathfrak{x}(\mathbf{x},t+\tau)$ 
is an operator-valued distribution (acting on $\mathcal{H}_S$) 
representing the abstract density $\hat{\mathfrak{x}}(\mathbf{x}, \tau)$ at time $t$.
Assuming that we only consider test functions $h$ with compact support in the time direction, we conclude that the operator 
$X(t)$ is localized in a time-slice, $I\times \mathbb{R}^{3}$, of finite width, where $I\equiv I_{X(t)}$ is a bounded interval of the 
time axis (assumed to contain the time $t$ in its interior), and $\mathbb{R}^{3}$ is physical space.\\

\textit{\underline{Pillar 2}: Heisenberg-picture dynamics of operators.}

Next, we recall how the time evolution of physical quantities in the \textit{Heisenberg picture} is described.
For this purpose we have to introduce the notion of an {\bf{isolated (physical) system}}.
An isolated system $S$ is one whose degrees of freedom have negligibly weak interactions with the 
degrees of freedom of its complement, $S^{c}$, i.e., with the rest of the Universe, during the period in 
time when the evolution of $S$ is monitored. (Yet, the state of $S\vee S^{c}$ can be entangled!) 
As discovered by \textit{Heisenberg} and Dirac, it is \textit{only} for isolated systems that we are 
able to formulate a general dynamical law for the time evolution of physical quantities. For simplicity, we also assume that the 
system $S$ is \textit{autonomous}. Then there exists a selfadjoint operator, $H=H^{*}$, 
acting on the Hilbert space $\mathcal{H}_S$, the \textit{Hamiltonian} of the system $S$, such that the operators representing an arbitrary physical quantity $\hat{X}\in \mathcal{O}_S$ at two different times, $t$ and $t'$, are unitarily conjugated to each other by the propagator generated by $H$, i.e.,
\begin{equation}\label{Heisenberg}
X(t') = e^{i(t'-t)H} \,X(t) \,e^{-i(t'-t)H}, \quad \text{for arbitrary times }\,\,\, t, t'\,,
\end{equation}
where $X(t)$ represents $\hat{X}$ at time $t$ (see Eq. \eqref{Rep}). This equation is 
commonly referred to as the \textit{Heisenberg equation}. It encapsulates the \textit{deterministic} law of time evolution of 
operators on $\mathcal{H}_S$ representing physical quantities in $\mathcal{O}_S$ characteristic of the system $S$. 
Notice that if $X(t)$ is localized in the time interval $I_{X(t)}$ then $X(t')$ is localized in the interval $I_{X(t')}= I_{X(t)} +(t'-t)$.

Eq. \eqref{Heisenberg} is usually extended to arbitrary bounded operators on $\mathcal{H}_S$:
\begin{equation}\label{Picture}
A(t) = e^{itH} \, A \, e^{-itH}\,, \quad \forall A \in B(\mathcal{H}_S)\,,
\end{equation}
for arbitrary times $t \in \mathbb{R}$. It is straightforward to extend Eqs.~\eqref{Heisenberg} and \eqref{Picture} to 
non-autonomous isolated systems, whose Hamiltonians are time-dependent.

\textit{Remark:} If there are substantial interactions between the degrees of freedom 
of $S$ and degrees of freedom describing the ``environment'', $S^{c}$, of $S$ the description of the time evolution 
of physical quantities characteristic of $S$, i.e., of operators representing elements of $\mathcal{O}_S$, can be 
arbitrarily complicated.\footnote{In this respect, classical mechanics is simpler than QM, because time evolution 
of a physical system in classical mechanics is always generated by a vector field on the state space of the system, 
albeit not necessarily a Hamiltonian vector field.}
(A description of the dynamics of systems interacting with their environment in terms of, for example, 
quantum Markov semi-groups generated by Lindblad operators is an approximation that, frequently, 
cannot be justified; although it is widely used.)\\

\textit{\underline{Pillar 3}: Expectation values of physical quantities in ``states.''}

In order to extract concrete information about the behavior of (an ensemble of identical) isolated physical systems, $S$, 
as described in Pillars 1 and 2, one has to be able to take expectation values of self-adjoint operators $X(t)$ on 
$\mathcal{H}_S$ representing physical quantities $\hat{X} \in \mathcal{O}_S$. For this purpose, one introduces 
some notion of \textit{``state''}. In non-relativistic quantum mechanics, ``states'' are usually taken to be 
\textit{density matrices} on $\mathcal{H}_S$, which are non-negative, trace-class operators, $\widetilde{\Omega}$, 
acting on $\mathcal{H}_S$ of trace 1, i.e.,
\begin{equation}\label{density matrix}
\widetilde{\Omega} = \widetilde{\Omega}^{*} \geq 0,\,\,\text {with }\,\, \text{tr}(\widetilde{\Omega}) =1\,.
\end{equation}
(In the following, we usually refer to these states as \textit{normal states}. For all possibly unfamiliar notions concerning  abstract functional analysis see, e.g.,  \cite{Lanford} and \cite{Takesaki}.) Pure states are given by orthogonal projections, $P=P^{*}=P^{2}$, of rank 1 corresponding to unit rays in $\mathcal{H}_S$.
The expectation, $\omega(X(t))$, of a physical quantity, $\hat{X}\in \mathcal{O}_S$, at time $t$ in a ``state'' given by a density matrix $\widetilde{\Omega}$ is defined by
\begin{equation}\label{Exp}
\omega(X(t)):= \text{tr}(\widetilde{\Omega}\cdot X(t))\,, 
\end{equation}
where $X(t)$ represents $\hat{X}$ at time $t$. Equation \eqref{Exp} is then extended to arbitrary bounded operators on $\mathcal{H}_S$, i.e., 
$$\omega(A):= \text{tr}(\widetilde{\Omega}\cdot A), \qquad \forall A \in B(\mathcal{H}_S)\,.$$
We will see shortly that this notion of ``state'' does not have any ontological meaning.\footnote{If a notion of ``state'' is 
supposed to have concrete physical (ontological) meaning then a ``state'' ought to be a functional that can only be evaluated 
on operators describing ``future potentialities'' (rather than on arbitrary operators acting on $\mathcal{H}_S$); 
see Sect. \ref{ETH}.}

It is common to claim that, in the Heisenberg picture, only operators evolve non-trivially in time, but states $\omega$ are \textit{time-independent}. One then usually goes on to claim that the Heisenberg picture is equivalent to the \textit{Schr\"odinger picture}, where physical quantities are \textit{time-independent} but ``states'' evolve in time according to the \textit{Schr\"odinger-Liouville equation}
\begin{equation}\label{Schroedinger}
\Omega(t)= e^{-i(t-t')H} \,\Omega(t') \, e^{i(t-t')H}\,, \quad \text{for arbitrary times }\,\, t, t'\,, \, \text{ with }\, \Omega(0)\equiv \widetilde{\Omega}\,.
\end{equation}
One then obviously has that 
$$\text{tr}\big(\widetilde{\Omega}\cdot A(t)\big)= \text{tr}\big(\Omega(t)\cdot A\big), \qquad \forall A \in B(\mathcal{H}_S)\,,$$
see Eqs. \eqref{Picture}, \eqref{Schroedinger}.

\textit{Remark:} In the following, we use a ``tilde'' to indicate that we refer to a density matrix in the Heisenberg picture, while we drop the ``tilde'' in the Schr\"odinger picture. We usually identify a density matrix in the Heisenberg picture with the corresponding density matrix in the Schr\"odinger picture at time $t=0$.

\subsection{Measurements and the Collapse Postulate}

\hspace{0.5cm}\textit{The concept of `measurement' becomes so fuzzy on reflection that it is quite surprising to have it appearing in physical theory ...} (John Stewart Bell)

We note that Eqs.~\eqref{Heisenberg}, \eqref{Picture} and \eqref{Schroedinger} are {\bf{linear}} and {\bf{deterministic}} 
evolution equations.
However, most physicists agree with the claim that the predictions of QM are {\bf{statistical}} (probabilistic). So, what is 
going on? Well, according to the Copenhagen Interpretation of QM, the Schr\"odinger-Liouville evolution of states given 
by Eq.~\eqref{Schroedinger} is interrupted whenever  \textit{``a measurement takes place,''} and in such a moment 
the Schr\"odinger-Liouville evolution is replaced by a non-linear change of state described -- at least heuristically  -- 
by the so-called \textit{collapse postulate}: If a physical quantity $\hat{X}$ is measured at some time $t$, with the outcome
 that it has a measured value $\xi \in \text{spec}(\hat{X})$, then the state, $\widetilde{\Omega}$, occupied by the system 
 $S$ right \textit{before} the measurement of $\hat{X}$ is carried out is supposed to be replaced by the state, 
 $\widetilde{\Omega}_{\xi, t}$, given by
\begin{equation}\label{collapse}
\widetilde{\Omega} \,\mapsto\,  \widetilde{\Omega}_{\xi, t} := \big[\text{tr}\big(\,\Pi_{\xi}(t)\big)\big]^{-1} \Pi_{\xi}(t)\, 
\widetilde{\Omega}\, \Pi_{\xi}(t)\,,
\end{equation}
right \textit{after} the measurement of $\hat{X}$, where $\Pi_{\xi}(t)$ is the spectral projection corresponding to the eigenvalue $\xi$ of the self-adjoint operator $X(t)$ representing $\hat{X}$ at time $t$. 

The question then arises what the precise quantum-mechanical law is that determines under what conditions a ``measurement'' is carried out and at what time the state-collapse \eqref{collapse} resulting from the measurement happens. 
Answering this question will amount to adding a \textit{``Fourth Pillar''} to the formulation of QM. Actually, the prescription 
in Eq.~\eqref{collapse} is \textit{at best} a reasonable heuristic recipe, but does obviously not have the status of a general law, 
as long as the notion of ``measurement'' remains totally vague and does not correspond to a well-defined operation in 
the mathematical formalism of QM\,!\footnote{This would not be too serious a problem if all we intended to do is to describe 
a \textit{single} measurement (at the ``end of time''), but gave up the requirement that the theory ought to also consistently 
describe \textit{repeated} measurements.} Before we will describe a precise general law governing the time evolution 
of states in QM (see Sects.~\ref{ETH}, \ref{HP}, \ref{Model}) we will argue that Schr\"odinger evolution does \textit{not} 
provide such a \mbox{law --} \textit{even} if all the experimental equipment used to perform measurements on a given 
system of interest is included in the quantum-mechanical description (so that the resulting \textit{total} system is \textit{isolated}). This will be discussed in the next section in the context of a gedanken experiment.
\newpage

\section{The Schr\"odinger equation does \textit{not} describe the time evolution of states in Quantum Mechanics}\label{Schroedinger evol}
\hspace{0.5cm}\textit{``I insist upon the view that `all is waves'.''} (Erwin Schr\"odinger, letter to J. L. Synge)\\

It is claimed, for valid reasons, that the Schr\"odinger equation describes the evolution of the states 
(``wave functions'') of so-called \textit{closed systems}.\footnote{We will explain what precisely this claim means in the next section.} 
 Unfortunately, though, it is impossible to gain any information on the behavior of closed systems, so that 
statements about them are statements about pure \textit{``potentialities''} which cannot be put to experimental tests. 
(But it is admitted that, to a good approximation, systems can behave as if they were closed over a long period of time.)

As most people will presumably agree upon, the Schr\"odinger equation does, however, 
\textit{not} describe the evolution of states of a physical system whenever a ``measurement'' or ``observation'' is 
performed on it that results in a \textit{``fact''} or \textit{``actuality''}. This insight is almost as old as Quantum Mechanics, 
in the form given to it by \textit{Born, Heisenberg} and \textit{Jordan}, and \textit{Dirac}. Its validity has been made 
plausible by deriving various paradoxes from the assumption that the time evolution of states in QM is \textit{always} 
described by a (linear) Schr\"odinger equation \textit{even} during periods when ``measurements/observations'' 
are made, as long as the total system we consider remains isolated.\footnote{Recall that an \textit{isolated system} 
is a physical system with the property that, for all practical purposes, interactions of the degrees of freedom of the 
system with those of its complement (the rest of the universe) can be neglected for the period of time during which 
the evolution of the system is monitored.}
It is worth stressing that the paradoxes do \textit{not} disappear \textit{even} if the measurement devices are included 
in a quantum-mechanical description of the system -- as \textit{has} to be done if we restrict our attention to 
\textit{isolated} systems -- and \textit{even} if linear Schr\"odinger evolution is amended by something like the 
\textit{Everett Interpretation} \cite{Everett} of QM. Among the more famous such paradoxes is the \textit{paradox 
of Wigner's friend} \cite{Wigner} and various more recent variants thereof; see \cite{Hardy, FR}. 

\subsection{Historical remarks}

\hspace{0.5cm}\textit{``[The probability wave] ... introduced something standing in the middle between the idea of an event and the actual event, a strange kind of physical reality just in the middle between possibility and reality.''} (Werner Heisenberg)

It turns out that issues concerning the meaning of states in QM and of their time evolution in the course of
measurements have been discussed, in a very clear way, by \textit{Heisenberg} in his 1955 lecture 
\cite{Heisenberg} at a conference on the occasion of the $75^{th}$ anniversary of \textit{Niels Bohr}. He points out 
that transitions from the ``potential'' to the ``factual'' (or ``actual'') are \textit{not} described by the Schr\"odinger equation 
and that, as a matter of fact, \textit{``this discontinuous change} [in the state of a system] \textit{is of course not 
contained in the mechanical equations of the system or of the ensemble characterizing the system; it corresponds [...] 
exactly to the ``reduction of the wave packets'' in quantum theory.''} He continues to describe his 
understanding of the \textit{Copenhagen Interpretation} of Quantum Mechanics. Along the way he discusses various 
basic quantum-mechanical phenomena; for example one that people nowadays call $\mathbf{decoherence}$, a phenomenon 
he claims is typically encountered when (macroscopic) measuring equipment is included in the description of the system. 
But he argues convincingly that decoherence will not solve the problem of how to 
understand the transition from the ``potential'' to the ``factual''. 
Referring to \textit{Bohr}, he argues that, for the functioning 
of some measuring equipment, it is crucial that there be a connection of the total system (system of interest to be 
measured or observed composed with the measuring equipment) to the ``outside world,'' because, as he says, 
\textit{``the behavior of the measuring equipment has to be registered as something factual.''} In other words, the total
system under consideration should be isolated and \textit{open}.
He then argues (without presenting compelling reasons) that, because of this connection to the outside 
world, \textit{``the state of the total system will now have to be  described by a mixed state,''} which 
cannot be understood to be the result of unitary (Schr\"odinger) time evolution.
He points out that a system completely disconnected from the outside world will only 
 have the character of the ``potential'', but not of the ``factual.'' He concludes by saying that, 
from the point of view of quantum theory, knowledge of the ``factual''\footnote{i.e., knowledge of ``actuality'' -- 
in Aristotelian jargon} always remains fundamentally \textit{incomplete knowledge}, and that, for this reason, 
\textit{``the statistical character of the microphysical laws will not disappear, anymore.''}
Although Heisenberg's analysis is impressively lucid, given that it was presented already
in 1955, it does have some weaknesses: 
\begin{itemize}
\item{While Heisenberg implicitly observes that, in Quantum Mechanics, the ontology is \textit{not} in 
the ``wave function'' (the state), he does not offer any precise ideas or opinions concerning the question of 
what the ontology is. To answer this question is one of our main concerns in this paper; see Sects.~\ref{ETH}, \ref{Model} 
and \ref{Conclusions}. In this connection, ideas described by \textit{Rudolf Haag} in \cite{Rudolf} are relevant and 
interesting.}
\item{While Heisenberg compellingly argues that, in QM, the evolution of states of physical systems subject to 
measurements, i.e., featuring events, is \textit{not} described by a Schr\"odinger equation and is fundamentally 
``statistical,'' he does not propose any precise \textit{``statistical law''} that describes the transitions from the 
``potential'' to the ``actual'' and the evolution of states of isolated open systems amenable to observation. 
To propose such a law is the main subject of our paper.}
\item{In 1955, the mathematical machinery was missing -- at least among elderly theoretical physicists -- 
to render the ideas Heisenberg describes in words in \cite{Heisenberg} mathematically precise. 
For example, his grasp of phenomena such as decoherence was intuitive and not expressed in 
mathematical formulae. This may explain why his remarkable paper on the interpretation of QM 
appears to have been unjustly ignored.}
\end{itemize}

\subsection{A concrete example showing that \textit{`not all is waves'} in QM}
\hspace{0.5cm} \textit{``Pick a flower on Earth and you move the farthest star.''} (Paul Adrien Maurice Dirac)

In the remainder of this section, we discuss an example of a concrete experiment that shows clearly that, in QM, the evolution 
of states of a physical system is \textit{not} described by a Schr\"odinger equation if the system features events that can be
observed from the outside. We follow arguments described in 
\cite{FFS, Schill}.\footnote{The paper \cite{FFS} is based on results that were first sketched around fifteen years ago.} \\

We consider an isolated physical system composed of two subsystems: 
\begin{itemize}
\item{a subsystem, $\mathfrak{L}$, consisting of a large detector, $\mathfrak{D}$, behind a spin filter (realized, for example, 
as a magnetized membrane of iron whose magnetization points in the vertical direction 
chosen to be the z-direction, as shown in \textit{Figure 1}, below), both described quantum-mechanically, composed with a 
particle $L$ with spin $1/2$ (an electron) hitting the spin filter with ``very high probability'' and either traversing 
it and then hitting the detector $\mathfrak{D}$, or being absorbed in an empty state of the spin filter 
(and possibly emitting a photon or phonon into the bulk of the filter);}
\item{a subsystem, $\mathfrak{R}$, consisting of a particle $R$ with spin $1/2$ (a silver atom),
which, with very high probability, passes through the magnetic field between two Stern-Gerlach magnets oriented 
in the direction of some unit vector $\vec{e} \in \mathbb{R}^{3}$ and then hits a detector $\mathfrak{D}_{+}$, provided
 its spin is parallel to $\vec{e}$, or a detector $\mathfrak{D}_{-}$ if its spin is anti-parallel to $\vec{e}$, 
 respectively; all of these ingredients constituting the subsystem $\mathfrak{R}$ (see \textit{Figure 1}).}
\end{itemize}
The initial state, $\psi_{L/R}$, of $L$ and $R$ is entangled: Their spin wave function forms 
a spin-singlet. One may imagine that, at time $t=0$, $\psi_{L/R}$ describes a negatively charged silver ion 
located in the central region between the spin filter and the Stern-Gerlach equipment, as indicated in \mbox{\textit{Figure 1}.} 
When a photon collides with the silver ion it kicks out an electron, and we assume that the resulting orbital wave function 
is such that the electron $L$ propagates into a cone, $\mathcal{C}'$, opening towards the spin filter in the left half-space, 
while the silver atom $R$ propagates into a cone, $\mathcal{C}$, opening towards the right half-space in the direction of the 
magnetic field between the Stern-Gerlach magnets -- with only very tiny chances to find the particles outside these 
two cones. 
\begin{center}
 $\mathcal{C'}$ \hspace{5.2cm} $\mathcal{C}$\\
\includegraphics[width=12cm]{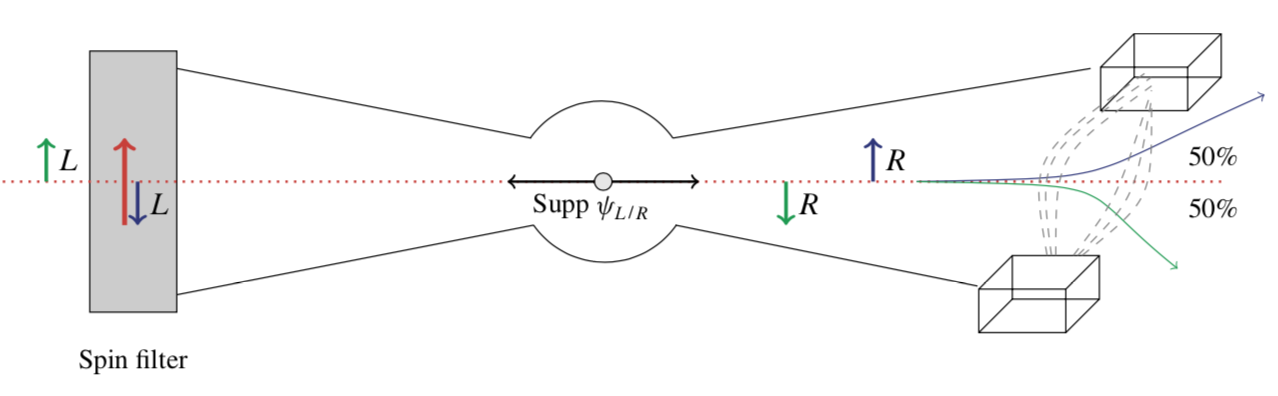}\\
\small{
$ \mathfrak{L}:=\lbrace$particle\, $L\, \vee $ spin filter\, $\vee$\, $\mathfrak{D} \rbrace$ \hspace{0.5cm} $\mathfrak{R}:=\{$particle $R \,\,\vee$ Stern-
Gerlach equipment $\vee  \,\, \mathfrak{D}_{+}, \,\,\mathfrak{D}_{-}\}$
}\\
\vspace{0.2cm}
\textit{Figure 1}
\end{center}
We temporarily assume that the state, $\Psi$, of the total system is prepared at time $t=0$ and 
then evolves according to some very complicated, but linear and deterministic Schr\"odinger equation. 
As shown in \cite{FFS} (and this is not entirely trivial!), assuming that we treat the system non-relativistically 
and neglect photons, with all interactions between different degrees of freedom of the total system of 
short range, the evolution of the subsystem $\mathfrak{R}$, consisting of the silver atom $R$, 
the Stern-Gerlach equipment and  the detectors $\mathfrak{D}_{+}$ and $\mathfrak{D}_{-}$, (\textit{after} the 
disintegration of the negative silver ion) is \textit{independent} of the evolution of the subsystem $\mathfrak{L}$, 
up to very tiny corrections. More precisely, when expectation values in the state $\Psi$ are taken the Heisenberg-picture 
time evolution of operators representing physical quantities referring to $\mathfrak{R}$ 
 (see \textit{Figure 1}) is approximately \textit{identical} to the one that 
would be used if the Hamiltonian of $\mathfrak{L}$, including interaction terms between $\mathfrak{L}$ and $\mathfrak{R}$, 
were set to 0. This has the following consequences: Let $\vec{S}_{R}\cdot \vec{e}$ denote the component of the spin 
operator of the silver atom along the direction of the unit vector $\vec{e}$ (see \cite{FFS}). The expectation value in 
the state $\Psi$ of this operator, evolved according to the Heisenberg-picture dynamics of the total system to some 
time $t$, is then well approximated by the expectation value of the same operator, but evolved to time $t$ according to the 
Heisenberg-picture dynamics determined by the Hamiltonian where the subsystem $\mathfrak{L}$ is \textit{absent}. 
But if the subsystem $\mathfrak{L}$ is absent the operator $\vec{S}_{R}\cdot \vec{e}$ is a conserved quantity, i.e., 
time-independent. Hence 
\begin{equation}\label{Bell R}
\langle e^{-itH} \Psi, (\vec{S}_{R}\cdot \vec{e})\, e^{-itH} \Psi \rangle \simeq \langle \Psi, (\vec{S}_{R}\cdot \vec{e})\, 
\Psi \rangle = 0,
\end{equation}
for times $t$ long enough for the electron $L$ to have reached the spin filter and the silver atom $R$ to have reached
the region where the detectors $\mathfrak{D}_{+}, \mathfrak{D}_{-}$ are installed; the operator $H$ being the 
Hamiltonian of the total system. We note that equality between the left side and the right side in \eqref{Bell R} 
may be violated by very tiny corrections that tend to 0 when the distance between the spin filter and the 
Stern-Gerlach equipment tends to $\infty$.

Next, we consider the subsystem $\mathfrak{L}$: It is plausible that the electron $L$ passes through the spin filter
 if its spin is aligned with the majority spin of electrons inside the spin filter, i.e., if its spin points in the positive z-direction. 
 The reason is that all electronic states (of low energy) localized inside the spin filter corresponding to a spin pointing in 
 the positive z-direction tend to be already occupied and, hence, cannot be occupied by $L$; as follows from the Pauli 
 exclusion principle. But if the spin of $L$ is anti-parallel to the majority spin of the spin filter then it finds plenty of empty 
 states localized inside the spin filter which it can hop into without violating the Pauli principle.
Thus, if the spin of $L$ were pointing in the positive z-direction it would eventually hit the detector $\mathfrak{D}$, which 
would then click, while if the spin of $L$ were pointing in the negative z-direction this particle would be absorbed by 
the spin filter, and the detector $\mathfrak{D}$ would remain silent.
The estimates contained in \cite{FFS} also show that 
\begin{equation}\label{Bell L}
\langle e^{-itH} \Psi, (\vec{S}_{L}\cdot \vec{e}_z)\, e^{-itH} \Psi \rangle \simeq \langle \Psi, (\vec{S}_{L}\cdot \vec{e}_z)\, 
\Psi \rangle = 0,
\end{equation}
for times $t$ long enough for the electron $L$ to have interacted with the spin filter and to have either been absorbed 
by the filter or hit the detector $\mathfrak{D}$. The predictions \eqref{Bell R} and \eqref{Bell L} are \textit{independent} 
of the relative orientation of $\vec{e}$  and the z-axis $\vec{e}_z$, and they remain true even if, for example, the direction of
$\vec{e}$ is suddenly changed, \textit{before} or \textit{after} the electron $L$ has passed through the spin filter, but 
\textit{before} the silver atom $R$ has reached the field region between the Stern-Gerlach magnets.

{\bf{Obviously}}, the predictions \eqref{Bell R} and \eqref{Bell L}, derived by solving the Schr\"odinger equation for 
the time evolution of the state $\Psi$ of the total system, do {\bf{not}} describe what is observed in experiments, i.e.,
they do not reflect the facts. For, in every successful experiment, the state of the total system occupied at large times 
$t$, when the experiment is completed, should exhibit the following features: 
The silver atom $R$ has passed through the magnetic field of the Stern-Gerlach magnets and has \textit{either} hit 
$\mathfrak{D}_{+}$ \textit{or} $\mathfrak{D}_{-}$, and the electron $L$ has \textit{either} been absorbed by the 
spin filter \textit{or} has hit the detector $\mathfrak{D}$. Of course, Quantum Mechanics does \textit{not} predict \textit{which} 
of these alternatives is realized in a single experiment. But it ought to predict that \textit{one} of these
precise alternatives (and not ``something in between'', i.e., not a coherent superposition of states corresponding 
to these alternatives) \textit{is} realized in every experiment! Alas, this is \textit{not} 
what one concludes by solving a deterministic linear Schr\"odinger equation, which would merely imply Eqs.~\eqref{Bell R} 
and \eqref{Bell L} -- and \textit{nothing more}. Furthermore, Quantum Mechanics must be given a form so as to predict
that if the experiment is repeated many times the emerging \textit{correlations} between the possible events described above
are given by the following well known formulae
\begin{equation}\label{probability}
\text{Prob}_{\Psi}\big\{L \text{ hits }\, \mathfrak{D}\,\,\, \& \, \,\,R \text{ hits } \,\mathfrak{D}_{-} \big\} = 
\frac{1}{4}\Big(1 + \vec{e}\cdot \vec{e}_z\Big)\,, \quad \text{etc.}
\end{equation}
This formula is claimed to correctly describe the correlations between measurements of components  of the spins of $L$ 
and of $R$ \textit{even} if, for example, the direction of $\vec{e}$ is suddenly changed \textit{before} the silver atom $R$ 
has traversed the magnetic field between the Stern-Gerlach magnets, but in such a way that the change of the direction 
of $\vec{e}$ cannot causally affect the fate of the electron $L$. This is what is often called the {\bf{``non-locality''}} of 
Quantum Mechanics (see, e.g., \cite{BGold}). It is not captured by Schr\"odinger evolution. Moreover, it is claimed that the formula applies 
\textit{independently} of whether $L$ hits $\mathfrak{D}$ \textit{before} or \textit{after} $R$ hits $\mathfrak{D}_{-}$.

Yet, in the \textit{Everett interpretation} of QM \cite{Everett} it is claimed that the time evolution of states of 
isolated physical systems is \textit{always} given by linear, unitary Schr\"odinger evolution (\textit{`all is waves'}), 
but that the universe can and will branch into a tree-like ``multiverse'' of alternative histories perceived as 
mutually exclusive. In experiments of the above kind the process of ``branching'' caused by subsystem $\mathfrak{R}$, 
i.e., by the Stern-Gerlach measurement of the spin of the silver atom is described as follows: In every branch 
of the ``multiverse'', $R$ will \textit{either} hit $\mathfrak{D}_{+}$ \textit{or} $\mathfrak{D}_{-}$ -- 
with these two alternatives actually existing in different branches (along different histories) of the 
``multiverse''.  Equation \eqref{Bell R} then tells us that the frequency of observing $R$ to hit the detector 
$\mathfrak{D}_{-}$ is equal to $\frac{1}{2}$ and, likewise, the frequency of observing 
$R$ to hit $\mathfrak{D}_{+}$ is equal to $\frac{1}{2}$, {\bf{independently}} of the orientation of $\vec{e}$ and 
{\bf{independently}} of what is happening in subsystem $\mathfrak{L}$, as long as processes in $\mathfrak{L}$ do 
\textit{not} cause any further branching. People will argue that if there is ``branching'' caused by subsystem 
$\mathfrak{R}$ then there will also be ``branching'' caused by subsystem $\mathfrak{L}$. However, a precise law 
telling us under what conditions ``branching'' takes place, how and \textit{when} it takes place and what (among 
a continuum of possibilities) the different ``branches'' correspond to and describe has, to our knowledge, never
been proposed. Some authors invoke \textit{decoherence} as an explanation of ``branching'', without spelling out 
what precisely they have in mind. Thus, as far as we understand it, the Everett interpretation of QM lacks precision 
and has remained vague. It should be superseded by an unambiguous, mathematically precise formalism that retains 
some of its attractive features. 

For the following considerations it is convenient to work in the Heisenberg picture, with operators depending on time 
in accordance with the usual Heisenberg equations of motion. Let $\pi_{L}^{+/-}(t)$ denote the projection onto states 
of the total system corresponding to the observation that, \textit{``at time $t$, the electron
$L$ has hit the detector $\mathfrak{D}$/$L$ has been absorbed in the spin filter''}, respectively; and let $\pi_{R}^{+/-}(t)$ 
denote the projection onto states corresponding to the observation that, \textit{``at time $t$, the silver atom $R$ has 
hit $\mathfrak{D}_{+}$/$R$ has hit $\mathfrak{D}_{-}$''}, respectively. 
We should ask how formulae such as \eqref{probability} can be derived from properties of the time evolution of states in 
QM. Let us imagine that, in a successful experiment, one has retrieved the information that the electron has passed the 
spin filter and has hit the detector $\mathfrak{D}$ at some time $t'$, which is interpreted as meaning that, at time $t'$, the 
spin of the electron is aligned with the $z$-axis, and that, at a time $t$, the silver atom has traversed the magnetic field 
and has hit the detector $\mathfrak{D}_{-}$, meaning that its spin is anti-parallel to the vector $\vec{e}$ at time $t$. 
According to the \textit{Copenhagen interpretation} of QM this would imply that, at times $t''\geq \text{max}(t',t)$, 
\textit{before} further events are recorded, the state of the system is \textit{not} given by $\Psi$, anymore, but it 
is given by 
\begin{equation}\label{collapse-1}
\pi_{L}^{+}(t') \cdot \pi_{R}^{-}(t)\, \Psi/ \Vert \pi_{L}^{+}(t') \cdot \pi_{R}^{-}(t) \,\Psi \Vert, 
\end{equation}
assuming w.l.o.g. that $t'\geq t$. Furthermore, the probability to reach this particular state in an experiment at a time 
$\geq t'$ is given by \textit{Born's Rule}:
\begin{equation}\label{BR}
 \text{Prob}_{\Psi}\big\{L \text{ hits }\, \mathfrak{D}\,\text{ at time }t'\,\, \&\,\, \, R \text{ hits } \,\mathfrak{D}_{-} \text{ at time } t \big\}= 
\Vert \pi_{L}^{+}(t') \cdot \pi_{R}^{-}(t) \,\Psi \Vert^{2}\,,
\end{equation}
and likewise for the other possible outcomes of this experiment. These probabilities should, however, \textit{not} depend on whether $t'>t$ or $t'<t$, and the \textit{conditional probabilities} 
$$\mathfrak{P}_{L, R}(+\vert \pm):=\text{Prob}_{\Psi}\big\{L \text { hits detector }\,\mathfrak{D} \,\vert \,\text{ given that }\, R\, \text{ hits }\, \mathfrak{D}_{\pm}\big\} $$
ought to have the property that
\begin{equation}\label{SR}
\sum_{\sigma=+, -} \mathfrak{P}_{L, R}(+\,\vert\, \sigma) \cdot  \text{Prob}_{\Psi}\big\{R \text{ hits }\, \mathfrak{D}_{\sigma}\big\} = \text{Prob}_{\Psi}\big\{L \text { hits } \, \mathfrak{D} \big\}\,,
\end{equation}
and similarly for $\mathfrak{P}_{L,R}(-\,\vert\,\sigma)$ (with ``$-$'' indicating that $L$ is absorbed in the spin filter).
For these sum rules to hold true -- no matter whether $t'>t$ or $t'<t$ -- it is sufficient that the following commutator vanish:
\begin{equation}\label{locality}
\big[  \pi_{L}^{\rho}(t') , \pi_{R}^{\sigma}(t) \big] = 0\,, \qquad \rho, \sigma = + \,\text{ or } \,-\,.
\end{equation}
This would follow from \textit{Einstein causality} in a model of the system satisfying the basic properties of local 
relativistic quantum theory (see \cite{Haag}), provided the operators $\pi_{R}^{\sigma}(t)$ and $ \pi_{L}^{\rho}(t')$ are 
localized in \textit{space-like separated} regions. This is precisely what, in relativistic Quantum Field Theory, is called 
{\bf{locality}} (of relativistic Quantum Theory)!

We conclude that, in order to reach agreement between theory and experiment (see, e.g., formula \eqref{probability}), 
one has to assume that, at certain times, the \textit{``wave function of the total system collapses''}, as described in 
Eqs. \eqref{collapse-1} and \eqref{BR}, with condition \eqref{SR} imposed for consistency. This collapse process 
is \textit{not} described by a Schr\"odinger equation (as emphasized by Heisenberg \cite{Heisenberg}). In 
standard QM, there is no law telling us whether a collapse is taking place and \textit{at which time} it takes place;  
the times $t$ and $t'$ in formulae \eqref{collapse-1} and \eqref{BR} are \textit{not} determined by the theory. 

We interpret formula \eqref{probability} and rules \eqref{collapse-1} through \eqref{SR} as indicating that the state 
of an isolated open system evolves \textit{stochastically} and \textit{non-linearly}.  A \textit{key concern} in this paper 
is to find the \textit{law} governing the stochastic evolution of states of an isolated physical system featuring events 
(such as, for example, measurements of physical quantities). This law ought to predict the possible states into which 
the state of the system may collapse when an event happens, and it ought to determine the probabilities of collapsing 
into specific states, as well as the approximate times at which the collapse will take place. Proposing such a law is 
the very subject of Sects.~\ref{ETH} and \ref{Model}.

\section{The Fourth Pillar of Quantum Mechanics -- summary of the $\mathbf{\textit{ETH}}$-Approach}\label{ETH}

\hspace{0.5cm}\textit{``Surely, after 62 years, we should have an exact formulation of some serious part of quantum mechanics. ... By `serious' I mean that some substantial fragment of physics should be covered.''} (John Stewart Bell)\\

In this section we endeavor to sketch a pragmatic formulation of QM, the \textit{ETH-Approach}, which is intended to 
eliminate those undesirable worldviews David Deutsch has been referring to in \cite{Deutsch}. In particular, it is intended
 to replace \textit{``interpretations''} of QM by a \textit{completion} of QM freed from puzzles such as the 
 \textit{``measurement problem.''} We view the $ETH$-Approach to QM as representing the \textit{Fourth Pillar} QM rests 
 upon. It is expected to provide stable foundations to the theory.
 
 The scope of this paper is limited to non-relativistic QM; see  
 \cite{BFS, Fr1}. The general ideas underlying the $ETH$-Approach can be extended to local relativistic 
 quantum theory, but the analysis becomes more subtle; for a beginning see \cite{Fr2}.
 
  \subsection{Algebras of potentialities and quantum probability measures}
 
 \hspace{0.5cm}\textit{``It is not the past that matters but the future.''} (Varun Ravikumar)
 
From Sect.~\ref{three pillars} we recall that operators, $X(t)$, representing physical quantities, $\hat{X}$, characteristic 
of a system $S$ at some time $t$  are self-adjoint operators, i.e., $X(t)=X(t)^{*}$, acting on a separable 
Hilbert space $\mathcal{H}_S$. As argued in Sect. \ref{three pillars}, one can associate a bounded interval 
$I_{X(t)}$ of the time axis containing $t$ with every such operator  $X(t)$; see Eqs. \eqref{Rep}, \eqref{phys quantity} 
and \eqref{Heisenberg}. It is natural to introduce algebras, $\mathcal{E}_I\,, I \subset \mathbb{R}$, as the algebras generated 
by arbitrary complex linear combinations of arbitrary products of operators $X(t)$ representing physical 
quantities $\hat{X} \in \mathcal{O}_S$ (see Eqs. \eqref{phys quantities}, \eqref{Rep}), with the property 
that $I_{X(t)} \subseteq I$, and of the identity, $\mathbf{1}$, on $\mathcal{H}_S$; (the identity $\mathbf{1}$ 
belongs to \textit{all} the algebras $\mathcal{E}_I$). We then define algebras $\mathcal{E}_{\geq t}$ as follows.
\begin{equation}\label{events}
\mathcal{E}_{\geq t} := \overline{\bigvee_{I \subset [t, \infty)} \mathcal{E}_I}\,,
\end{equation}
where (to be specific) the closure on the right side is taken in the topology of weak convergence on $\mathcal{H}_S$. 
The algebra $\mathcal{E}_{\geq t}$ is called the \textit{algebra of all potentialities at times $\geq t$}. (It is a von Neumann algebra.) It follows directly from the definition that 
\begin{equation}\label{inclusion}
\mathcal{E}_{\geq t'} \subseteq \mathcal{E}_{\geq t}\,, \quad \forall t' > t\,.
\end{equation}
We also define $\mathcal{E}$ to be the norm-closure of the algebra generated by 
$\big\{ \mathcal{E}_{\geq t} \big\}_{t\in \mathbb{R}}$; ($\mathcal{E}$ is the algebra of all potentialities in the history of the system $S$).\\

\textit{Remark}: Let $S$ be an isolated autonomous system with Hamiltonian $H$, and let $t'>t$. Then Eq. \eqref{Heisenberg} 
for the time evolution of operators in the Heisenberg picture implies that
\begin{equation}\label{automorphism}
\mathcal{E}_{\geq t'}= \big\{e^{i(t'-t)H} X\, e^{-i(t'-t)H} \,\vert\, X\in \mathcal{E}_{\geq t}\big\}\,\subseteq \mathcal{E}_{\geq t}\,,
\end{equation}
i.e., time evolution of operators in the Heisenberg picture by an amount $t'-t>0$ determines a \textit{$^{*}$endomorphism} 
of $\mathcal{E}_{\geq t}$ whose image is the algebra $\mathcal{E}_{\geq t'}$. It turns out that this important feature 
distinguishes the $ETH$-Approach  from various rather vague schemes based on the observation that time evolution 
of a system may entangle its degrees of freedom with those of an unobserved or unobservable environment. (We will 
come back to this point in Sect.~\ref{Conclusions})\\

\underline{Definition 1}: (Potentialities)

A \textit{potential event} or \textit{potentiality} associated with the system $S$ that might set in at time $t$ (i.e., is localized
at times $\geq t$) is given by a partition of unity by orthogonal projections on $\mathcal{H}_S$,
$$\big\{ \pi_{\xi}\, \vert \,\xi \in \mathfrak{X} \big\} \subset \mathcal{E}_{\geq t}\,,$$
where $\mathfrak{X}$ is a countable set, with the following properties:
\begin{align}\label{potentialty}
\quad \pi_{\xi}= \pi_{\xi}^{*} \in \mathcal{E}_{\geq t}, \quad \forall \xi \in &\mathfrak{X}\,,\quad
\pi_{\xi}\cdot\pi_{\eta} = \delta_{\xi \eta}\, \pi_{\xi} \,, \quad \forall\, \xi, \eta \text{ in }\, \mathfrak{X}\,, \quad
\text{ and } \quad \sum_{\xi \in \mathfrak{X}} \pi_{\xi} = \mathbf{1}\,.\quad \square
\end{align}

Let $\mathcal{P}_{\geq t}$ be the lattice of all orthogonal projections in $\mathcal{E}_{\geq t}$. \\

\underline{Definition 2}: (Quantum probabilities)

A \textit{quantum probability measure} on the potentialities localized at times  $\geq t$ is a map 
$\mu\,: \mathcal{P}_{\geq t} \rightarrow [0, 1]$ with the following properties:
\begin{enumerate}
\item[(i)]{ $0 \leq \mu(\pi) \leq 1\,, \forall\, \pi \in \mathcal{P}_{\geq t},$ \,\,with \,\,\,$\mu(0)=0\,$ \, and \, $\mu(\mathbf{1})=1\,$;}
\item[(ii)]{ $\mu\big(\sum_{\xi \in \mathfrak{X}_0} \pi_{\xi}\big) = 
\sum_{\xi \in \mathfrak{X}_0} \mu\big(\pi_{\xi}\big)\,,$
\,for an arbitrary potentiality $\big\{ \pi_{\xi}\, \vert\, \xi \in \mathfrak{X} \big\} \subset \mathcal{E}_{\geq t}$ and 
an arbitrary subset $\mathfrak{X}_0 \subseteq \mathfrak{X}$\,.
\hspace{11cm}$\square$}
\end{enumerate}

The following generalization of \textit{Gleason}'s theorem \cite{Gleason} follows directly from a general theorem due to  
\textit{Maeda} \cite{Maeda}.\\

{\bf{Theorem~1}}: (Gleason-Maeda) \textit{We assume that the algebras $\mathcal{E}_{\geq t}$ do not have any direct 
summand given by the algebra of all complex $2\times 2$ matrices. Then every probability measure, $\mu$, on the 
potentialities setting in at time $t$ is given by a normal state, $\omega_{\mu}$, on the von Neumann algebra 
$\mathcal{E}_{\geq t}$, with \\ 
\vspace{0.1cm}\hspace{5.9cm}
$ \omega_{\mu}(\pi) = \mu(\pi)\,,\quad  \forall \pi \in \mathcal{P}_{\geq t}\,. \hspace{4cm} \square$
}\\

\noindent
(A normal state on a von Neumann algebra $\mathfrak{M}$ is defined to be a positive linear 
functional, $\omega$, on $\mathfrak{M}$ continuous in the weak topology and normalized such that $\omega(\mathbf{1})=1$.)\\

\underline{Remarks}: 
\begin{enumerate}
\item{If $\big\{ \pi_{\xi}\, \vert \,\xi \in \mathfrak{X} \big\} \subset \mathcal{E}_{\geq t}$ is a potential event localized at times 
$\geq t$ then $\big\{ e^{isH}\pi_{\xi} e^{-isH} \,\vert\, \xi \in \mathfrak{X} \big\} \in \mathcal{E}_{\geq (t+s)} $ is a 
potential event localized at times $\geq (t+s)$. Here $H$ is the Hamiltonian of the system.} 
\item{For autonomous systems $S$ with \textit{finitely many degrees of freedom}, the algebras $\mathcal{E}_{\geq t}$ coincide with 
the algebra $B(\mathcal{H}_S)$ of all bounded operators on $\mathcal{H}_S$ and, hence, are \textit{independent} of $t$. It 
turns out that, for such systems, it is \textit{impossible} to introduce a non-trivial notion of \textit{events actually happening} 
(actualities) at some time $t$ or later, \textit{and the so-called {\bf{measurement problem}} cannot be solved by only considering such systems}. 
The situation is radically different if one considers systems for which the inclusions in \eqref{inclusion} are \textit{strict}, which 
can happen for systems with infinitely many degrees of freedom including ones describing \textit{massless modes}, such as photons and gravitons, 
that can escape to infinity, at the limiting speed, $c$, without being detected; see Sect.~\ref{HP}.}
\end{enumerate}

 \underline{Definition 3}: (Closed systems)
 
 A physical system $S$ is said to be a \textit{closed system} iff the algebras $\mathcal{E}_{\geq t}$ of all potentialities 
 setting in at time $t$ are \textit{independent} of $t$, for all times $t\in \mathbb{R}$ 
 (i.e., equality holds in \eqref{inclusion}, for all times $t$ and $t'$). Closed systems 
 have the same defects as systems with finitely many degrees of freedom: The 
 measurement problem cannot be solved for such systems.
 
 \subsection{The Principle of Diminishing Potentialities}

\hspace{0.5cm}\textit{``Indeed, it is evident that the mere passage of time itself is destructive rather than 
\mbox{generative ..., } because change is primarily a `passing away.' ''}  (Aristotle, Physics )

In order to introduce a good notion of \textit{events actually setting in at some time $t$} (for short: actualities) and to clarify 
how such events can be recorded in \textit{projective measurements}, we require the following \\

\underline{Principle of Diminishing Potentialities} $(PDP)$: An isolated system $S$ featuring actualities, i.e., events that set in 
at some (finite) time, has the property that
\begin{equation}\label{PDP}
 \hspace{4.2cm}  \mathcal{E}_{\geq t'} \subsetneqq \mathcal{E}_{\geq t} \subsetneqq \mathcal{E}\,, \quad \text{whenever}\,\,\,\, t'>t\,. \qquad\, (PDP) \hspace{2.4cm} \square
 \end{equation}

This principle has been introduced and analyzed in \cite{F-Schub, BFS, Fr1, Fr2}.\footnote{In earlier work, see \cite{BFS}, $(PDP)$ has 
been called \textit{Loss of Access to Information (LAI)}} 
Our main concern in this paper is to describe a concrete family of models satisfying the Principle of Diminishing Potentialities; 
see Sects.~\ref{HP} and \ref{Model}. In the models of Sect.~\ref{HP}, the algebras $\mathcal{E}_{\geq t}$ are associated with 
physical quantities localized inside future light cones nested inside one another, and $(PDP)$ turns out to be a 
consequence of the existence of massless modes, e.g., photons, whose dynamics satisfies locality or Einstein 
causality (\textit{Huygens' Principle}; see \cite{Buchholz} for a general analysis).
The models studied in Sect.~\ref{Model} arise in the limit of the speed of light tending to $\infty$. We expect that, in 
\textit{all} models of \textit{non-relativistic} Quantum Mechanics with a \textit{continuous time}, $(PDP)$ only holds 
if the spectrum of the Hamiltonian, $H$, of the system is \textit{unbounded above and below}; see Sect. \ref{Conclusions}. 
However, as argued in \cite{Fr2} using important results in \cite{Buchholz}, in \textit{local relativistic quantum theory}, 
$(PDP)$ is compatible with the spectrum condition 
$H\geq 0$.\\

{\bf{Description of isolated open systems.}}\\

An \textit{isolated, but open system}, $S$, is described, quantum-mechanically, in terms of a \textit{co-filtration} (i.e., a decreasing filtration)
$\lbrace \mathcal{E}_{\geq t} \rbrace_{t \in \mathbb{R}}$ (or $\lbrace \mathcal{E}_{\geq t} \rbrace_{t \in \mathbb{Z}}$, in case $time$ is assumed to be discrete, see Sects.~\ref{HP} and \ref{Model}), of von 
Neumann algebras, $\mathcal{E}_{\geq t}$, satisfying $(PDP)$, all represented on a common Hilbert space $\mathcal{H}_S$, 
whose lattices of projections describe potentialities. \\

Let $\omega$ be a state occupied by $S$, as introduced in Eq.~\eqref{Exp} of Sect.~\ref{three pillars}; (see \cite{Fr-Schub} for an analysis of how to prepare a system in a specific state). The state, 
$\omega_t$, of $S$ at time $t$,
\begin{equation}\label{state-rest}
\omega_{\,t}(X):= \omega(X), \qquad \forall X \in \mathcal{E}_{\geq t},
\end{equation}
is defined to be the restriction of the state $\omega$ to the algebra $\mathcal{E}_{\geq t}$. By the Gleason-Maeda theorem, 
$\omega_{\,t}$  
corresponds to a quantum probability measure on $\mathcal{P}_{\geq t}$. The state $\omega_{\,t}$ of $S$ 
at some time $t$, as defined in \eqref{state-rest}, will usually be a \textit{mixed} state  \textit{even} if $\omega$ 
is a pure state on $\mathcal{E}$. This is a consequence of $(PDP)$, Eq.~\eqref{PDP}, and \textit{entanglement}; (see Sects.~\ref{HP} and \ref{Model} for explicit examples). We should then clarify what we mean by saying that
$\omega_{\,t}$ is a \textit{mixed state}, and what the implications of this property for the appearance 
of \textit{``actual events''} are.
We begin by formulating a criterion enabling us to decide whether an actual event sets in at time 
$t$.\footnote{This criterion is inspired by the desire to rescue as many of the more reasonable features of the \textit{Copenhagen interpretation} of quantum mechanics as possible.} More precisely, the criterion formulated below 
enables us to decide whether, given a state, $\omega_t$, on the algebra $\mathcal{E}_{\geq t}$, there exists a potential 
event localized at times $\geq t$ that describes an \textit{actual event} setting in at time $t$.\\

\underline{Definition 4}: (The centralizer of a state) \label{defcentralizer}

Given a $^{*}$-algebra $\mathcal{A}$ and a state $\omega$ on $\mathcal{A}$, the centralizer, 
$\mathcal{C}_{\omega}(\mathcal{A})$, of the state $\omega$ is the subalgebra of all operators $Y \in \mathcal{A}$ 
with the property that 
$$\omega([Y, X]) =0, \qquad \forall X \in \mathcal{A},$$
i.e.,
$$\hspace{2.8cm}\mathcal{C}_{\omega}(\mathcal{A}):=\left\{Y\in \mathcal{A}\, \vert \,\,\omega([Y, X]) =0, \,\,\forall X \in \mathcal{A}\right\} . 
\hspace{2.8cm} \square$$

We note in passing that the state $\omega$ defines a finite (normalized) trace on its centralizer 
$\mathcal{C}_{\omega}(\mathcal{A})$. This enables one to classify all those von Neumann algebras that can arise as 
centralizers of normal states on von Neumann algebras.\\

\underline{Definition 5}: (The center of the centralizer)

The \textit{center} of the centralizer $\mathcal{C}_{\omega}(\mathcal{A})$ of the state $\omega$, denoted by 
$\mathcal{Z}_{\omega}(\mathcal{A})$, is the abelian subalgebra of $\mathcal{C}_{\omega}(\mathcal{A})$ consisting
of all operators that commute with \textit{all} other operators in 
$\mathcal{C}_{\omega}(\mathcal{A})$, i.e.,
$$
\hspace{2.5cm}\mathcal{Z}_{\omega}(\mathcal{A}):= \big\{ Y \in \mathcal{C}_{\omega}(\mathcal{A})| \,\,[Y,X]=0\,,\, \forall X \in 
\mathcal{C}_{\omega}(\mathcal{A}) \big\}. \hspace{2.3cm} \square
$$

 We note that the center, $\mathcal{Z}(\mathcal{A})$, of an algebra $\mathcal{A}$ is contained in 
 $\mathcal{Z}_{\omega}(\mathcal{A})$, for all states $\omega$ on $\mathcal{A}$. After these preparations, 
 we define \textit{actual events/actualities} in a system $S$ in the following way.\\
 
 \underline{Definition 6}: (Actual events/actualities) \label{ETHevent}
 
Let $S$ be an isolated open system described by a co-filtration $\big\{ \mathcal{E}_{\geq t} \big\}_{t \in \mathbb{R}}$ of von 
Neumann algebras. Given a state $\omega_t$ on the algebra $\mathcal{E}_{\geq t}$, an \textit{actual event} corresponding 
to a potential event described  by a partition of unity 
$\big\{ \pi_{\xi}\,\vert\, \xi \in \mathfrak{X} \big\} \subset \mathcal{E}_{\geq t}$\, is setting in
 at time $t$ iff $\mathcal{Z}_{\omega_t}(\mathcal{E}_{\geq t})$ is non-trivial,\footnote{The 
 algebra $\mathcal{Z}_{\omega_{t}}(\mathcal{E}_{\geq t})$ is an \textit{abelian} von Neumann algebra.
On a separable Hilbert space, it is generated by a single self-adjoint operator $X$ (an ``observable''), 
whose spectral projections yield the projections $\lbrace \pi_{\xi}\,\vert \, \xi \in \mathfrak{X} \rbrace$
describing a potential event. This is the motivation behind \eqref{act-event}.}
\begin{equation}\label{act-event}
\big\{ \pi_{\xi}\,\vert \,\xi \in \mathfrak{X} \big\} \,\,\text{generates   }\,\, \mathcal{Z}_{\omega_{\,t}}\big(\mathcal{E}_{\geq t}\big),
\end{equation}
and the \textit{Born probabilities}
\begin{equation}\label{Born}
\omega_{\,t}(\pi_{\xi_{j}}) \,\,\text{ are \textit{strictly positive}}\,, \,\, \text{ for points }\,\xi_j \in \mathfrak{X}, \,\, j=1,2, \dots, n\,, 
\end{equation}
for some~$n\geq 2$. \hspace{12cm}$\square$

According to this definition, for a \textit{potential event} 
$\big\{\pi_{\xi}\,\vert\, \xi \in \mathfrak{X}\big\}\subset \mathcal{E}_{\geq t}$ to be an
\textit{actual event} featured by the system $S$ during a time period contained in $[t, \infty)$ it is apparently necessary 
and sufficient that the projections $\big\{\pi_{\xi}\,\vert\, \xi \in \mathfrak{X}\big\}$ generate the center, 
$\mathcal{Z}_{\omega_t}(\mathcal{E}_{\geq t})$, of the centralizer of the state $\omega_t$ on the algebra 
$\mathcal{E}_{\geq t}$.\\ 

Next, we propose to analyze the consequences of the statement that, in some isolated open system $S$, 
an actual event or actuality sets in at time $t$. 

\subsection{Actual events and the Collapse Postulate}

\hspace{0.5cm}\textit{``Every experiment destroys some of the knowledge of the system which was obtained by previous experiments.''} (Werner Heisenberg)

Let $\omega_{\,t}$ be the state of~$S$ right \textit{before} time $t$. Let us assume that an actual event
 $\big\{ \pi_{\xi} \,|\, \xi \in \mathfrak{X} \big\}$ generating $\mathcal{Z}_{\omega_t}(\mathcal{E}_{\geq t})$ sets in (i.e., begins to unfold) at time $t$. This implies that
 \begin{equation}\label{incoh superpos}
 \omega_{\,t}(A)= \sum_{\xi \in \mathfrak{X}} \omega_{\,t}(\pi_{\xi} \,A\, \pi_{\xi}), \qquad \forall A \in \mathcal{E}_{\geq t}\,,
 \end{equation}
 i.e., $\omega_{\,t}$ is an \textit{incoherent superposition} of states in the range of the projections 
 $\pi_{\xi}, \xi \in \mathfrak{X}$; (no off-diagonal terms appear on the right side of Eq. \eqref{incoh superpos}). In other words, the quantum probability measure determined by $\omega_{\,t}$ on the potentialities, $\mathcal{P}_{\geq t}$, at times $\geq t$ is a convex combination of quantum probability measures indexed by the points $\xi \in \mathfrak{X}$ that label the projections of the actual event setting in at time $t$. (In this precise sense, $\omega_t$ is a \textit{mixture} indexed by the points of $\mathfrak{X}$.)\\
 
\underline{Pillar 4}:  In the $ETH$-Approach to QM, the following axiom (see \cite{Fr1}) is required in order to complete the mathematical formulation of Quantum Mechanics:

{\bf{Axiom CP}} (Collapse Postulate).\label{collapseaxiom} \textit{Let $S$ be an isolated open system satisfying $(PDP)$. Let
$\omega_t$ be the state on the algebra $\mathcal{E}_{\geq t}$ right \textit{before} time $t$. Let 
$\big\{ \pi_{\xi} \,|\, \xi \in \mathfrak{X} \big\} \subset \mathcal{Z}_{\omega_t}(\mathcal{E}_{\geq t})$ be the actual event (actuality) setting in at time $t$. Then the state on $\mathcal{E}_{\geq t}$ occupied by $S$ right \textit{after} the event has set in is given by 
$$
\omega_{\,t, \xi_{*}}(\cdot):=[\omega_{\,t}(\pi_{\xi_{*}})]^{-1}\,\omega_{\,t}\big(\pi_{\xi_{*}} (\cdot) \pi_{\xi_{*}}\big) \,,
$$
for some point $\xi_{*} \in \mathfrak{X}$ with $\omega_{\,t}(\pi_{\xi_{*}})>0$. The probability for the system $S$ to be found in the state $\omega_{\,t,\xi_{*}}$ right \textit{after} time $t$ is given by {\bf{Born's Rule}}, i.e., by}
\begin{equation}\label{Born Rule}
\hspace{4.9cm}\text{\rm{prob}}\{\xi_{*}, t\} = \omega_{\,t}(\pi_{\xi_{*}}). \hspace{4.5cm} \square
\vspace{0.2cm}
\end{equation}

\underline{Remark}: In \textit{local relativistic quantum theory}, this axiom has to be replaced by a somewhat similar, though
rather more subtle one, which incorporates the structure of the bundle of light cones in space-time and Einstein causality in an 
interesting way; see \cite{Fr2}. We will return to this topic in forthcoming work. 

The $ETH$-Approach to QM yields the following picture of the \textit{dynamics of states} in Quantum Mechanics: 
The evolution of states of an isolated open system $S$ featuring events, in the sense of Definition 6 stated above, 
is determined by a (continuous-time) \textit{stochastic branching process}, whose state space is referred to 
as the {\em{non-commutative spectrum}}, $\mathfrak{Z}_{S}$, of $S$ (see~\cite{Fr1}). Assuming that all the algebras 
$\mathcal{E}_{\geq t}$ are isomorphic to one specific (universal) von Neumann algebra, denoted by $\mathcal{N}$,\footnote{This is the case in the models considered in Sect.~\ref{Model} and in relativistic Quantum Electrodynmics  \cite{Buchholz}} the non-commutative spectrum, $\mathfrak{Z}_{S}$, of $S$ is defined by
\begin{equation}\label{NCspect}
\mathfrak{Z}_{S}:= \bigcup_{\omega} \Big(\,\omega, \mathcal{Z}_{\omega}(\mathcal{N})\Big)\,, 
\end{equation}
where the union over $\omega$ is a disjoint union, and $\omega$ ranges over \textit{all} states of $S$ of physical interest.\footnote{``States of physical interest'' are normal states a concrete system can actually be prepared in. Here we leave this notion a little vagues; but see \cite{Haag, Fr-Schub}.}
Born's Rule \eqref{Born Rule}, together with ~\eqref{act-event}, then specifies the \textit{branching probabilities} 
of the process. (See also Sects.~\ref{Model} and \ref{Conclusions}.)\\

\underline{Remarks}:
\begin{enumerate}
\item{Here is an explanation of the meaning of the name ``$ETH$-Approach'': ``$E$'' stands for ``events'', 
``$T$'' for ``trees'' -- referring to the tree-like structure of the space of all actualities an isolated physical system 
could in principle encounter in the course of its evolution --, and ``$H$'' stands for ``histories'' -- 
referring to the \textit{actual trajectory} of states occupied by the system in the course of its evolution.}
\item{{\bf{Axiom CP}} (the Collapse Postulate) formulated above, in combination with Eqs.~\eqref{incoh superpos} 
and \eqref{Born Rule}, is reminiscent of the collapse postulate in the Copenhagen interpretation of QM. But, thanks to the 
Principle of Diminishing Potentialities $(PDP)$, its status in the $ETH$-Approach to QM is not only logically consistent, but perfectly natural (i.e., not \textit{ad hoc}). That the \textit{non-commutative spectrum} $\mathfrak{Z}_S$ plays a very important role in an analysis of the time evolution of states becomes strikingly clear in local relativistic quantum theory; see \cite{Fr2}.

One might argue that  $(PDP)$ and the Collapse Postulate provide a mathematically precise version of the 
\textit{Many-Worlds Interpretation} of QM. However, in the $ETH$-Approach, there is no reason, whatsoever, to 
imagine that many alternative worlds actually \textit{exist}!}
\item{In \cite{Fr1, Fr2}, we have explained in which way the occurrence of an actual event may correspond 
to the measurement of a physical quantity. In general, there are plenty of actualities happening that cannot be related 
to the measurement of a previously specified physical quantity, and there is \textit{no role} to be played by ``observers'' (let alone their consciousness) in the 
$ETH$-Approach to QM.

An analysis of observations and measurements in QM and of how measurements are used to record 
events in the $ETH$-Approach has been presented in \cite{Fr1, Fr2} (see, in particular, Sect. 5 of \cite{Fr2}). 
It will not be repeated here. Suffice it to say that  an \textit{actuality} setting in at time $t$, 
described by a partition of unity $\big\{\pi_{\xi}\,\vert\, \xi \in \mathfrak{X}\big\}\subset \mathcal{E}_{\geq t}$, corresponds 
to measuring a physical quantitiy $\hat{X}\in \mathcal{O}_S$ (see Eq.~\eqref{phys quantities}) iff the projections 
$\big\{\pi_{\xi}\,\vert\, \xi \in \mathfrak{X}\big\}$ can be well approximated (in the norm on the linear space $B(\mathcal{H}_S)$ 
given by the scalar product induced by the state $\omega_t$ -- see \cite{Fr1, Fr2}) by \textit{spectral projections} of the self-adjoint operator $X(t')\in \mathcal{E}_{\geq t}$ representing $\hat{X}$ at a time $t'\geq t$, with $t' \approx t$. This will be clarified in Sect.~\ref{Model} in the context of simple models.
} 
\item{We hope that the stochastic branching processes on the non-commutative spectra of isolated open systems
derived from $(PDP)$ and the Collapse Postulate, along with Eqs. \eqref{act-event}, \eqref{Born} and \eqref{Born Rule}, 
will attract the interest of mathematicians. A beginning of an analysis of the simplest such processes is presented in Sect.~\ref{Model}, which is the section containing the main new message of our paper.}
\end{enumerate}

\section{Huygens' Principle and the Principle of Diminishing Potentialities}\label{HP}
\hspace{0.5cm}\textit{``... principles are tested by inferences which are derivable from them. The nature of the subject permits of no other treatment.''} (Christiaan Huygens)\\

In this section, we explain why and how the existence of massless modes (photons or gravitons) in an isolated physical 
system implies the validity of the Principle of Diminishing Potentialities $(PDP)$. We introduce a class of models of 
isolated systems for which this claim can be verified explicitly. The material discussed in this section also serves 
to motivate the models studied in Sect.~\ref{Model}.

We consider an isolated system, $S$, consisting of a very heavy (actually infinitely heavy) atom interacting with the 
quantized electromagnetic field. The atom is located in a compact region centered at the 
origin, $\mathbf{x}=\mathbf{0}$, of physical space $\mathbb{R}^{3}$. Gravitational effects are neglected. 
Points in space-time, $\mathbb{M}^{4}$, (Minkowski space) are denoted by $x=(x^{0}\equiv c\, t, \mathbf{x})$, where  
$c$ is the speed of light, and $\mathbf{x}=(x^1,x^2,x^3)$ is a point in physical space $\mathbb{R}^{3}$. Let $F_{\mu\nu}(x) \equiv F_{\mu\nu}(\mathbf{x}, t)$ be the field tensor of the 
quantized free electromagnetic field, which is an operator-valued distribution on Minkowski space.
 If $\big\{h^{\mu \nu}(\mathbf{x},t)\,\vert\,\mu, \nu = 0,1,2,3\big\}$ are real-valued test functions on $\mathbb{M}^{4}$ then 
\begin{equation}\label{field tensor}
F(h):= \int_{\mathbb{M}^{4}} d^{4}x \,F_{\mu \nu}(\mathbf{x},t) \, h^{\mu \nu}(\mathbf{x},t)
\end{equation}
turns out to be a self-adjoint operator on the \textit{Fock space}, $\mathcal{F}$, of the free electromagnetic field; 
(see, e.g., \cite{Derezinski}). We may then consider bounded functions of the operators $F(h)$, which are bounded 
operators on $\mathcal{F}$.

In a space-time description, the system $S$ is located, at time $t$, in a compact region of $\mathbb{M}^{4}$ centered at 
$x=(c\, t, \mathbf{x}= \mathbf{0})$. Let $V^{+}_t$  be the (closure of the interior of the) forward light cone with vertex 
at the space-time point $(c\, t, \mathbf{0})$, and, likewise, let $V^{-}_t$ be the 
backward light cone with vertex at  $(c\, t, \mathbf{0})$. For $t<t'$, we define the (space-time) \textit{diamond} 
$D_{t,t'}$ by setting
\begin{equation}\label{diamond}
D_{t,t'}:= V^{+}_t \cap V^{-}_{t'} \,.
\end{equation} 
We define $\mathcal{A}_{[t,t']}$ to be the von Neumann algebra generated by all bounded functions of the operators
$F(h)$, where $\big\{h^{\mu\nu}(x)\,\vert \, \mu, \nu =0, 1, 2, 3, x \in \mathbb{M}^{4} \big\}$ are real-valued test functions 
on $\mathbb{M}^{4}$ with 
support in the diamond $D_{t, t'}$. For an arbitrary time $t\in\mathbb{R}$, we define the algebra $\mathcal{A}_{\geq t}$ to be the von Neumann algebra 
generated by all the algebras $\mathcal{A}_{[t,t']}, t'>t$; i.e.,
\begin{equation}\label{em algebras}
\mathcal{A}_{\geq t} := \overline{\bigvee_{\mathbb{R}\ni t'>t} \mathcal{A}_{[t,t']}}\,.
\end{equation}
We suppose that, besides the quantized electromagnetic field, $S$ has ``internal'' degrees of freedom corresponding to excited states of the atom. Transitions between these states are described by operators acting on a (possibly only finite-dimensional) Hilbert space $\mathfrak{h}_S$.
 The Hilbert space of pure state vectors of $S$ is thus given by
$$\mathcal{H}_S = \mathcal{F} \otimes \mathfrak{h}_S\,.$$
Operators representing physical quantities characteristic of $S$ generate algebras of operators acting on 
$\mathcal{H}_S$ that are defined as follows:
\begin{align}\label{alg of events}
\mathcal{D}_{[t,t']}^{(0)}:= &\mathcal{A}_{[t,t']} \otimes \mathbf{1}\vert_{\mathfrak{h}_S}\,,\quad \mathcal{E}_{[t,t']}^{(0)} := \mathcal{A}_{[t,t']} \otimes B(\mathfrak{h}_S)\,,\quad \text{for arbitrary  }\, t<t' ,\nonumber \\
&\mathcal{E}_{\geq t}^{(0)}:= \mathcal{A}_{\geq t} \otimes B(\mathfrak{h}_S), \quad \text{for all times  } \, t , \quad  \mathcal{E}:= B(\mathcal{H}_S)\,.
\end{align}
The algebra $\mathcal{E}_{\geq t}^{(0)}$ may be interpreted as the \textit{algebra of all potentialities at times $\geq t$}, as long as the internal degrees of freedom of the atom are not coupled to the electromagnetic field; (see Eqs. \eqref{events}, \eqref{inclusion}, \mbox{Sect. \ref{ETH}).}

If $\mathfrak{M} \subset B(\mathcal{H}_S)$ is a (von Neumann) algebra of operators acting on $\mathcal{H}_S$ then
$\mathfrak{M}\,'$ is defined to be the (von Neumann) algebra of \textit{all} bounded operators on $\mathcal{H}_S$ commuting with \textit{all}
 operators in $\mathfrak{M}$. The following lemma is an easy exercise (see \cite{Buchholz} for more general results and interesting applications).\\
 
{\bf{Lemma 2}}: (``Huygens' Principle'') 
\begin{equation}\label{Huygens}
\big[ \mathcal{E}_{\geq t'}^{(0)}\big]' \cap \mathcal{E}_{\geq t}^{(0)} = \mathcal{D}_{[t,t']}^{(0)}
\end{equation}

{\bf{Sketch of proof}}: Obviously, every operator in $\big[ \mathcal{E}_{\geq t'}^{(0)}\big]'$ must commute with all operators of 
the form $\mathbf{1}\vert_{\mathcal{F}} \otimes C, \, C \in B(\mathcal{H}_S)$, i.e., it must have the form 
$ A\otimes \mathbf{1}\vert_{\mathfrak{h}_S}$, where $A$ is a bounded function of the electromagnetic field 
operators. If $A$ belongs to $\mathcal{A}_{\geq t}$, as it must if $A\otimes \mathbf{1}$ belongs to 
$\mathcal{E}_{\geq t}^{(0)}$, then $A$ is a bounded function of the field operators $F(h)$, 
for test functions $\big\{h^{\mu\nu}\big\}$ 
supported in $V^{+}_{t}$. For the free electromagnetic field, a field operator $F(h)$ commutes with \textit{all} field operators 
$F(g)$ affiliated\footnote{An unbounded self-adjoint operator $A$ is affiliated with a von Neumann algebra $\mathfrak{M}$ if and only if all bounded functions (spectral projections) of $A$ belong to $\mathfrak{M}$.} with $\mathcal{A}_{\geq t'}$, i.e., with supp$(g^{\mu\nu}) \subseteq V^{+}_{t'}$, for $\mu, \nu= 0,1,2,3$, if and only if 
supp$(h^{\mu\nu}) \subseteq V^{-}_{t'}$, $\forall\, \mu, \nu$. This is a straightforward consequence of the fact 
that the commutator distributions (which, for the free electromagnetic field, are all proportional to a c-number distribution 
on $\mathbb{M}^{4}$ solving the wave equation) satisfy
\begin{equation}\label{locality}
\big[F_{\mu\nu}(x), F_{\rho \sigma}(y)\big] = 0, \qquad \text{unless }\,\,(x-y)^{2} =0\,,
\end{equation}
i.e., unless $x-y$ is a lightlike vector. This is called \textit{Huygens' Principle} in quantum field theory. Thus, 
if $F(h)$ is affiliated with $\mathcal{A}_{\geq t}$ and commutes with all operators in $\mathcal{A}_{\geq t'}$ then 
\begin{equation}\label{support}
\text{supp}(h^{\mu\nu}) \subseteq V^{+}_t \cap V^{-}_{t'} = D_{[t,t']}\,, \qquad \forall\, \mu, \nu\,.
\end{equation}
Bounded functions of the operators $F(h)\otimes \mathbf{1}\vert_{\mathfrak{h}_S}$, with $\big\{h^{\mu \nu}\big\}$ a real-valued test function satisfying \eqref{support}, generate the algebra $\mathcal{D}_{[t,t']}^{(0)}$, and it follows from Eq.~\eqref{locality} and results in \cite{DrFr} that they commute with all operators in $\mathcal{E}_{\geq t'}^{(0)}$. This completes the proof of the lemma. \hspace{3.4cm}$\square$\\

We note that the algebras $\mathcal{D}_{[t,t']}^{(0)}$ are infinite-dimensional.\footnote{Actually, these algebras are 
von Neumann algebras (factors) of type $III_{1}$; see \cite{Buchholz, Takesaki}} Lemma 2 and this last fact show 
that a very strong form of the \textit{Principle of Diminishing Potentialities},
$$
\mathcal{E}_{\geq t'}^{(0)} \subsetneqq \mathcal{E}_{\geq t}^{(0)} \subsetneqq \mathcal{E}^{(0)}= B(\mathcal{H}_S), \quad
 \text{whenever }\,\,\, t< t' < \infty\,,
$$
holds for the system considered here, as long as the atom is not coupled to the quantized radiation field.

So far, only the free electromagnetic field has played a role in our discussion. The dynamics of the ``internal'' 
degrees of freedom of the system $S$ described by operators acting on $\mathfrak{h}_S$ will be specified next, and we also describe how these degrees of freedom of $S$ are coupled to the electromagnetic field. Let $H_0$ denote the usual Hamiltonian of the free electromagnetic field; ($H_0\geq 0$ is a self-adjoint operator on $\mathcal{H}_S$, see, e.g., \cite{Derezinski}). We imagine that the internal degrees of freedom of the atom are driven periodically in time with some period $T$, which (w.l.o.g.) we may set to 1. We choose a unitary operator $U\equiv U_{1}\in \mathcal{E}_{[0,1]}^{(0)}$ and set
\begin{align}\label{S-dyn}
U_k:= e^{i(k-1)H_0} &U_{1} e^{-i(k-1)H_0}\,, \quad k=1,2,3, \dots\,, \quad U(n):=\prod_{k=1}^{n} U_{k} \,,\, \text{ with }\, U(0):=\mathbf{1}\,,\nonumber \\
 & \Gamma:= e^{-iH_0}U_1, \qquad \Gamma^{n} = e^{-inH_0} U(n)\,,\quad n=0,1, 2,3,\dots \,.
\end{align}
Notice that 
$$\Gamma^{n}\cdot \Gamma^{m}= \Gamma^{n+m}\,, \qquad \forall\,\, n, m \quad \text{belonging to}\quad \mathbb{Z}_{+}:=\{0,1,2,3, \dots \}\,.$$
We consider $\big\{ \Gamma^{n} \big\}_{n\in \mathbb{Z}_{+}}$ to be the propagator of $S$, and conjugation with this 
propagator describes the time evolution of operators in the Heisenberg picture when the atom is coupled to 
the electromagnetic field. At time $t=0$, $S$ is prepared in a state $\omega_0$ given, for example, by
\begin{equation}\label{vacuum}
\omega_{0}(X):= \text{tr}(\big[\,\vert 0\rangle \langle 0\vert \otimes \widetilde{\Omega}\,\big] \, X)\,,
\end{equation}
where $X$ is an arbitrary operator in $\mathcal{E}_{\geq 0}:= \mathcal{E}_{\geq 0}^{(0)}$, $\vert 0\rangle$ is the vacuum vector in Fock space 
$\mathcal{F}$, and $\widetilde{\Omega}$ is some density matrix on $\mathfrak{h}_S$. To simplify our analysis, we 
henceforth regard time as discrete, $t=n \in \mathbb{Z}_{+}$, and monitor the evolution of $S$ only for non-negative times. 

In accordance with the definition of the propagator $ \big\{\Gamma^{n}\big\}_{n\in \mathbb{Z}_{+}}$ in Eq. \eqref{S-dyn}, we define algebras of potentialities, $\mathcal{E}_{\geq n}$, at times $\geq n$, with $n=0,1,2,\dots$, for the \textit{interacting system} as follows:
\begin{equation}\label{pot}
\mathcal{E}:= \mathcal{E}_{\geq 0}^{(0)}\,, \quad \mathcal{E}_{\geq n} := \Big\{\Gamma^{-n}\, X\, \Gamma^{n}\, 
\vert \,X \in \mathcal{E} \Big\}\,, \quad \text{with }\,\, \Gamma^{-1}= \Gamma^{*}.
\end{equation}
Apparently, the algebras ${\mathcal{E}}_{\geq n}$ are related to the algebras $\mathcal{E}_{\geq n}^{(0)}$ by unitary conjugation, the unitary operator being $U(n)$, i.e., every operator $Z\in {\mathcal{E}}_{\geq n}$ is of the form
$$ Z= U(n)^{*} X U(n),\qquad \text{for some   }\,\,  X\in \mathcal{E}_{\geq n}^{(0)}\,,$$
for arbitrary $n\in \mathbb{Z}_{+}$; see \eqref{S-dyn}. For arbitrary $n<n'$, we define the algebra $\mathcal{D}_{[n,n']}$ to be given by conjugating all operators in $\mathcal{D}_{[n,n']}^{(0)}$ by the unitary operator $U(n')$. With these definitions, Lemma 2 implies that, for the interacting system, too, the Principle of Diminishing Potentialities, see Eq. \eqref{PDP}, holds for all times $n$ and $n'$, with $ n'>n\geq 0$, and we have that
$$\big[\mathcal{E}_{\geq n'}\big]' \cap \mathcal{E}_{\geq n} \simeq \mathcal{D}_{[n,n']}\,.$$
It is important to note that 
\begin{equation}\label{nesting}
\Gamma^{\,n'-n}\,\mathcal{E}_{\geq n}\, \Gamma^{\,n-n'} = \mathcal{E}_{\geq n'} \subsetneqq \mathcal{E}_{\geq n}\,, \quad \text{ for }\,\, n'>n\,.
\end{equation}
If time is restricted to the integers ($t=n\in \mathbb{Z}_+$) the prescriptions in Definition 6, Eqs.~\eqref{act-event} 
and \eqref{Born}, and in {\bf{Axiom CP}}, Eq.~\eqref{Born Rule}, of Sect.~\ref{ETH} determine a \textit{stochastic branching 
process} on the non-commutative spectrum $\mathfrak{Z}_S:= \bigcup_{\omega} \big(\,\omega, \mathcal{Z}_{\omega}
(\mathcal{N})\,\big)$, where $\mathcal{N} \simeq \mathcal{E}_{\geq 0}^{(0)}$, see \eqref{NCspect}, and $\omega$ ranges 
over all states given by density matrices $\widetilde{\Omega}$ on $\mathcal{H}_S$. It is quite subtle to describe this 
process explicitly, because it is \textit{``non-Markovian''}, i.e., it has \textit{memory.} To understand this 
claim we choose two intergers $\ell$ and $m$, with $\ell < m$, and consider the algebras 
$$\mathcal{A}_{\leq \ell} :=\overline{\bigvee_{\{k\vert k< \ell\}} \mathcal{A}_{[k,\ell]}}\,, \quad \text{and }\,\, \mathcal{A}_{\geq m}\,,$$
see Eq.~\eqref{em algebras}.
 The arguments used in the proof of Lemma 2 show that 
$$\mathcal{A}_{\leq \ell} \subset \big(\mathcal{A}_{\geq m}\big)'\,,$$
 whenever $\ell < m$. A product state, $\varphi$, on $B(\mathcal{F})$ is a state with the property
 \begin{equation}\label{correlation}
\varphi(A \cdot B) = \varphi(A)\cdot \varphi(B)\, \qquad \forall\, \,A \in \mathcal{A}_{\leq \ell}, \, \,\,
\forall\,\, B \in \mathcal{A}_{\geq m}\,.
 \end{equation}

 It turns out that there are \textit{no product states} (among all states of physical interest \cite{Haag}); in particular, the state $\omega_0$ is \textit{not} a product state. It is this fact that implies that, in the model considered here, there are memory effects in the stochastic evolution of states determined by the law encoded in Eqs.~\eqref{act-event}, \eqref{Born} and \eqref{Born Rule} of Sect.~\ref{ETH}.
 It should be stressed, furthermore, that this evolution is typically non-linear.
  \textit{Only} for propagators $\big\{\Gamma^{n}\big\}_{n\in \mathbb{Z}_+}$ that do \textit{not} create any entanglement 
  between the internal degrees of freedom of the atom and the electromagnetic field, the evolution of states of  $S$ 
  in the \textit{Schr\"odinger picture} is described by the usual unitary Schr\"odinger evolution.
 
 We will not present a more detailed study of the models described above in the present paper.
  
 In the next section we simplify matters by studying models that arise in the limit
 \begin{equation}\label{limit}
 c\, \rightarrow \, \infty\,.
 \end{equation}
In this limit, the diamonds $D_{t,t'}$ open up to entire \textit{time slices}; i.e., 
$$D_{t,t'}= \big\{(\tau, \mathbf{x})\, \vert \, t \leq \tau \leq t', \mathbf{x} \in \mathbf{R}^{3}\big\}\,.$$
If we require -- as we will -- that Lemma 2 continues to hold in the limit considered in \eqref{limit} the algebras 
$\mathcal{D}_{[t,t']}$ must be contained in the commutant of all the algebras $\mathcal{D}_{[s,s']}$, 
whenever $t<t'\leq s<s'$. It turns out that the resulting limiting models have plenty of \textit{product states} of physical interest 
satisfying \eqref{correlation}. The price to be payed is that the Hamiltonian, $H_{0}^{c=\infty}$, of the caricature of the 
radiation field corresponding to the limit \eqref{limit} is \textit{unbounded} above \textit{and} below. (It has the same spectrum as the usual momentum operator.) This appears to be a general feature of models of non-relativistic QM satisfying the Principle of Diminishing Potentialities.

\section{Simple models illustrating the $\mathbf{\textit{ETH}}$-Approach to Quantum Mechanics}\label{Model}

\hspace{0.5cm}\textit{``One of the characteristic traits of collapse models is radiation emission from any charged particle induced by the noise causing the collapse of the wave function.''} (Introduction to \cite{ABDZ})\\

In this section we study a system $S$ that is composed of a very heavy atom (as in Sect.~\ref{HP}) coupled to a 
caricature of the quantized electromagnetic field, hereafter called \textit{``R-field''} (for ``radiation field''), obtained 
in the limit \eqref{limit} of the speed of light, $c$, tending to infinity. We introduce and analyze a class of simple 
models of $S$ that supply examples of explicit dynamical laws governing the time evolution of states of $S$ and 
hence illustrate the \textit{$ETH$-Approach} to non-relativistic QM described in Sect.~\ref{ETH}. In order to be able 
to carry out detailed calculations without appealing to high-brow mathematics, we adopt some \textit{drastic simplifications}:
\begin{enumerate}
\item[(1)]{Time $t$ is discrete: the time axis $\mathbb{R}$ is replaced by $\mathbb{Z}$.}
\item[(2)]{ The Hilbert space, $\mathfrak{h}_S$, of the internal degrees of freedom of the atom is finite-dimensional,
\begin{equation}\label{atom}
\mathfrak{h}_S \simeq \mathbb{C}^{M}, \qquad \text{for some }\,\, M<\infty\,.
\end{equation}
Physical quantities of the system $S$ referring to the atom are described by self-adjoint operators on $\mathfrak{h}_S$, 
i.e., by hermitian $M\times M$ matrices. General states of the atom are described by density matrices, $\widetilde{\Omega}$, 
acting on $\mathfrak{h}_S$.} 
\item[(3)]{The algebra generated by functions of the $R$-field localized in the time slice $[n,n+1]$ is chosen to be finite-dimensional, namely
\begin{equation}\label{algebra at fixed time}
\mathcal{A}_{n}\equiv \mathcal{A}_{[n,n+1]} \simeq \mathbb{M}_{N}(\mathbb{C}), \qquad \text{for some  }\,\, N<\infty\,,
\end{equation}
i.e., by all $N\times N$ complex matrices acting on the $N$-dimensional Hilbert space \mbox{$\mathcal{H}_n \simeq 
\mathbb{C}^{N}$.} \footnote{It is easy to extend our arguments to models with $\mathcal{H}_n$ 
infinite-dimensional; but this does not render the analysis more interesting.} We choose an orthonormal basis 
$\big\{\phi_j\big\}_{j=0}^{N-1}$ in $\mathbb{C}^{N}$. The \textit{``vacuum vector''} of the $R$-field is then defined by
\begin{equation}\label{vacuum}
\vert 0 \rangle \equiv \Phi_{\underline{0}}:= \bigotimes_{n\in \mathbb{Z}} \phi_{k_n =0}.
\end{equation} 
The interpretation of the vacuum vector $\Phi_{\underline{0}}$ is that it is the state where no modes of the $R$-field 
are excited. Taking this vector as a so-called \textit{reference vector}, the Hilbert space, $\mathfrak{F}_{S}$, 
of pure state vectors of the $R$-field is defined as follows: 
Let $\mathcal{S}_{fin}$ be the set of infinite sequences 
$\underline{k}:= \big\{k_n\big\}_{n\in \mathbb{Z}}$ with the property that $k_n =0$, except for finitely many values 
of $n$. To a sequence $\underline{k}\in \mathcal{S}_{fin}$ we associate a (tensor-) product vector $\Phi_{\underline{k}}$ by setting
\begin{equation}\label{CONS}
\Phi_{\underline{k}}:= \bigotimes_{n\in \mathbb{Z}} \phi_{k_n}\qquad \underline{k} \in \mathcal{S}_{fin}\,.
\end{equation}
Every such vector belongs to the non-separable Hilbert space 
$\mathfrak{F}_{\infty}:=\bigotimes_{n\in \mathbb{Z}} \mathcal{H}_n$. Let 
$\mathfrak{D}$ denote the linear subspace of $\mathfrak{F}_{\infty}$ consisting of all \textit{finite} linear combinations of 
vectors $\Phi_{\underline{k}}$, with $\underline{k}\in \mathcal{S}_{fin}$; $\mathfrak{D}$ is equipped with a scalar product 
$\langle \cdot, \cdot \rangle$ determined by
\begin{equation}\label{scalar prod}
\langle \Phi_{\underline{k}}, \Phi_{\underline{k}'}\rangle = \prod_{n \in \mathbb{Z}} \delta_{k_n, k'_n}\,,
\end{equation}
which is extended to $\mathfrak{D}\times \mathfrak{D}$ anti-linearly in the first argument and linearly in the second argument. 
The Hilbert space $\mathfrak{F}_{S}$ is then defined to be the completion of the space $\mathfrak{D}$ in the norm given by 
$$\Vert \Psi \Vert:= \sqrt{\langle \Psi, \Psi \rangle}, \qquad \Psi \in \mathfrak{D}\,.$$
The vectors $\big\{\Phi_{\underline{k}}\,\vert\, \underline{k}\in \mathcal{S}_{fin}\big\}$ form an orthonormal basis in 
$\mathfrak{F}_S$, hence $\mathfrak{F}_S$ is separable.

\underline{Remark}: It is interesting to consider other choices of reference vectors, $\Phi$, in the definition of the Hilbert space 
$\mathfrak{F}_S$. Examples will be given at the end of this section.

The total Hilbert space of the system $S$ is defined by
\begin{equation}
\mathcal{H}_S:= \mathfrak{F}_{S}\otimes \mathfrak{h}_S\,.
\end{equation}
}
\end{enumerate}

\subsection{Choice of algebras of operators representing potential events}

\hspace{0.5cm}\textit{``With respect to the property of Direction, the `possible' is called the Future and the `actualized' 
the Past.''} (Oswald Spengler)

Next, we introduce algebras of operators appropriate to describe the system $S$, assuming first that the atom 
is \textit{not} coupled to the $R$-field. Our definitions are similar to those introduced in Sect. \ref{HP}.\\

\underline{Definition 7}: (Algebras generated by the $R$-field)\\
For every integer $j$, we embed the algebra $\mathcal{A}_j\simeq \mathbb{M}_{N}(\mathbb{C})$ into the algebras $B(\mathfrak{F}_S)$ and 
$B(\mathfrak{F}_{\infty})$ by taking the tensor product of any operator $F\vert_{\mathcal{H}_j} \in \mathcal{A}_j$ with the identity operators 
$\mathbf{1}\vert_{\mathcal{H}_{\ell}}$ on all spaces $\mathcal{H}_{\ell}$, with $\ell\not= j$. The resulting algebras are again 
denoted by $\mathcal{A}_j$.
We then set 
$$\mathcal{A}_{[n,n']}:= \bigotimes_{j=n}^{n'-1} \mathcal{A}_{j}\,, \quad \text{for  }\,\, n'> n\,, \qquad 
\mathcal{A}_{\geq n}:= \overline{\bigvee_{n'>n} \mathcal{A}_{[n,n']}}\,.\footnote{the closure being taken in the weak topology of $B(\mathfrak{F}_S)$.}$$
As in Eq.~\eqref{events}, we introduce the following algebras.
\begin{align}\label{event algebras}
\hspace{1.6cm}\mathcal{D}_{[n,n']}^{(0)}:= \mathcal{A}_{[n,n']} \otimes \mathbf{1}\vert_{\mathfrak{h}_{S}}&,\quad 
\mathcal{E}_{[n,n']}^{(0)}:= \mathcal{A}_{[n,n']}\otimes B(\mathfrak{h}_{S}), \quad n<n'\,,\nonumber \\
&\mathcal{E}_{\geq n}^{(0)}:=\mathcal{A}_{\geq n} \otimes B(\mathfrak{h}_S)\,. \hspace{4.6cm}\square
\end{align}
One immediately checks that 
\begin{equation}\label{PDP-model}
\big[\mathcal{E}_{\geq n'}^{(0)}\big]' \cap \mathcal{E}_{\geq n}^{(0)} = \mathcal{D}_{[n,n']}^{(0)}, \qquad \forall n'>n\,.
\end{equation}
This imples that the \textit{Principle of Diminishing Potentialities (PDP)} -- see Eq.~\eqref{PDP} of Sect.~\ref{ETH} -- 
holds in this model, as long as the atom is not coupled to the $R$-field, yet.\\

We observe that the algebras $\mathcal{A}_{[n,n']}$ are isomorphic to $B(\bigotimes_{j=n}^{n'-1}\mathcal{H}_{j})$, for 
arbitrary  $n<n'$, and that the states
\begin{equation}\label{product states}
\varphi_{\underline{k}}(F):= \langle \Phi_{\underline{k}}, F\, \Phi_{\underline{k}} \rangle\,, \,\, \qquad F \in B(\mathfrak{F}_S)\,,
\end{equation}
are \textit{product states}, in the sense of Eq.~\eqref{correlation}, Sect.~\ref{HP}. For these reasons, there are no memory effects in the time evolution of states of the $R$-field, taken to be density matrices on $\mathfrak{F}_S$, as time $t$ tends to $\infty$. This will be shown in Subsect. 6.3.

In the following, we will only monitor the time evolution of states of $S$ for times  $t\geq t_{in}$, where $t_{in}$ 
is an \textit{initial time} that, in the following, we set to 0, and we then choose the algebra of physical quantities 
characteristic of $S$ to be given by $\mathcal{E}= \mathcal{E}_{\geq 0}^{(0)}$. 

\subsection{Time evolution in the Heisenberg picture}

\hspace{0.5cm} \textit{``The constant element in physics, since Newton, is not a configuration or a geometrical form, but a law of dynamics.''} (Werner Heisenberg)

Next, we describe the Heisenberg-picture time evolution of operators representing physical quantities characteristic 
of the system $S$; (see Eq.~\eqref{Heisenberg}, Sect.~\ref{three pillars}). The ``free'' time evolution of the $R$-field, before it is coupled to the atom, is given in terms of
a ``shift operator,'' $\mathfrak{S}$, on $\mathfrak{F}_S$:
\begin{equation}\label{shift map}
\big(\mathfrak{S}\Phi\big)_{\underline{k}}:= \Phi_{\sigma(\underline{k})} = \bigotimes_{n\in \mathbb{Z}} \phi_{\sigma(\underline{k})_{n}}, \quad \text{with  }\,\,\, \sigma(\underline{k})_n := k_{n+1}\,, \,\forall\, n\in \mathbb{Z}.
\end{equation}
The definition of the operator $\mathfrak{S}$ is extended to the domain $\mathfrak{D}$ (dense in $\mathfrak{F}_S$) by 
linearity. It is obviously unitary on $\mathfrak{D}$ and hence extends to all of $\mathfrak{F}_S$. We observe that 
$\mathfrak{S}$ leaves the vacuum (reference) vector $\vert 0 \rangle \equiv \Phi_{\underline{0}} \in \mathfrak{F}_S$ 
(see Eq.~\eqref{vacuum}) invariant. The shift operator $\mathfrak{S}$ is the analogue of the operator $e^{-iH_0}$ 
considered in Sect.~\ref{HP}. If time were chosen to be continuous then the generator of time evolution of the $R$-field 
would be an operator unitarily equivalent to a standard \textit{momentum operator} and hence would be unbounded 
from above and from below.

For an arbitrary operator $F\in \mathcal{A}_{[n,n']}$, we set
$$F(t):=\big[\mathfrak{S}^{*}\big]^{t}\,F\, \mathfrak{S}^{t}, \qquad t \in \mathbb{Z}\,,$$
and find that $F(t) \in \mathcal{A}_{[n+t,n'+t]}$, for arbitrary $n<n'$. Let $V$ be a unitary matrix on the atomic Hilbert space 
$\mathfrak{h}_S = \mathbb{C}^{M}$ describing the propagator of the atom by one time step. Before the atom is coupled to the $R$-field the Heisenberg-picture time evolution of bounded operators, $C$, on $\mathcal{H}_S$ is given by conjugation with the unitary propagator 
$\big\{\Gamma_{0}^{\,t}\big\}_{t\in \mathbb{Z}}$ on $\mathcal{H}_S$, 
\begin{equation*}
C(t):=\big[\Gamma_{0}^{*}\big]^{t}\, C \, \Gamma_{0}^{\,t} = \Gamma_{0}^{-t}\,C\,\Gamma_{0}^{t}\,, \qquad t\in \mathbb{Z}\,, 
\end{equation*}
where
\begin{equation}\label{propagator}
\Gamma_{0} := \mathfrak{S} \otimes V
\end{equation}
is a unitary operator on $\mathcal{H}_S$. We observe that 
$$
 \mathcal{E}_{\geq (n+t)}^{(0)}=\Big\{\Gamma_{0}^{-t}\, X\, \Gamma_{0}^{\,t} \, \vert \,X \in  \mathcal{E}_{\geq n}^{(0)}\Big\} ,\qquad \forall \, n, t \text{ in }\, \mathbb{Z}\,.
$$
For $t\geq 0,\, \mathcal{E}_{\geq (n+t)}^{(0)}$ is contained in $\mathcal{E}_{\geq n}^{(0)}$\,, i.e., Heisenberg-picture
time evolution by a time step $t\geq 0$ of operators on $\mathcal{H}_S$ defines a $^{*}$\textit{endomorphism} of
 $\mathcal{E}_{\geq n}^{(0)}$, for arbitrary $n \in \mathbb{Z}$. For a strictly positive $t$, $\mathcal{E}_{\geq (n+t)}^{(0)}$ is properly 
 contained in $\mathcal{E}_{\geq n}^{(0)}$, and $(PDP)$ holds. \\
 
 Next, we introduce {\bf{interactions}} between the atom and the $R$-field. As announced, we only monitor the 
 evolution of states of $S$ for non-negative times, $t\in \mathbb{Z}_{+}$. We choose a \textit{unitary} operator 
 $U$ in the algebra $B(\mathbb{C}^{N}) \otimes B(\mathfrak{h}_S)$ and define $U_1$ to be the corresponding 
 operator in the algebra \mbox{$\mathcal{A}_{0} \otimes B(\mathfrak{h}_S)$;} see Eq.~\eqref{algebra at fixed time}. We define
 \begin{align}\label{emission op}
 U_k &:= \Gamma_{0}^{1-k} \,U_1\, \Gamma_{0}^{k-1}, \,\,\, k=1,2,3,\dots ,\nonumber \\
 U(n):=  U_{n} \cdots& U_{1} \,, \,\,n=1,2, 3,\dots,\quad \Gamma:= \Gamma_{0} U_1,\,\,\, \Gamma^{-1}= \Gamma^{*}\,,
 \end{align}
 see Eq.~\eqref{S-dyn} for comparison. Notice that 
 $$U_k \in \mathcal{A}_{k-1}\otimes B(\mathfrak{h}_S)\,, \quad k=1,2,3,\dots $$
 It is straightforward to verify that the operators $U(n), n=1,2, \dots,$ and $\Gamma$ are unitary, and that
 \begin{equation}\label{semi-group}
 \Gamma^{\,n}=\Gamma_{0}^{\,n} U(n)\,,\,\,\, \forall \,\,\, n \in \mathbb{Z}_{+}\,. 
 \end{equation}
 We interpret $\big\{\Gamma^{\,n}\big\}_{n\in \mathbb{Z}_{+}}$ (with $\Gamma^{0}:= \mathbf{1}$) as the unitary propagator of the system $S$ in the presence of interactions between the idealized atom and the $R$-field.\\

\underline{Definition 8}:  Event algebras $\mathcal{E}_{\geq n}$ (with $n\geq 0$ henceforth) of the \textit{interacting} system are defined by setting
\begin{equation}\label{event alg}
\mathcal{E}:= \mathcal{E}_{\geq 0}^{(0)}\,, \quad \mathcal{E}_{\geq n} := \Big\{\Gamma^{-n}\, X\, \Gamma^{\,n}\, \vert\, X \in \mathcal{E} \Big\}\,,\quad n=0,1,2,\dots\,, \text{\,\,with\,\,\,\,}\Gamma^{-1}= \Gamma^{*}\,. \quad\,\, \square
\end{equation}
See also Sect.~\ref{HP}, Eq.~\eqref{pot}.

From now on, we identify physical quantities, $\hat{X}$, characteristic of the system $S$ with operators $X\equiv X(0) \in 
\mathcal{E}$ given by a sum of operators of the form $F\otimes C$, where $F\in \mathcal{A}_{\geq 0}$ and $C$ is 
an $M\times M$ matrix on $\mathfrak{h}_S$; and we set $X(t)= \Gamma^{-t}\,X\,\Gamma^{\,t}\,, t\in \mathbb{Z}.$ We then have that
$$\mathcal{E}_{\geq n'} = \Gamma^{-t}\, \mathcal{E}_{\geq n} \, \Gamma^{\,t} \subseteq \mathcal{E}_{\geq n}\,, \quad \text{for  }\,\, t= n'-n\geq 0\,,$$
i.e., time evolution by a time step $t\geq 0$ is given by a $^{*}$endomorphism from $\mathcal{E}_{\geq n}$ to 
$\mathcal{E}_{\geq (n+t)}\subseteq \mathcal{E}_{\geq n}$.

We observe that every operator $Z\in \mathcal{E}_{\geq n}$ is of the form
\begin{equation}\label{unit equiv}
Z= U(n)^{*}\, Y \,U(n),\quad \text{for some }\,\, Y\in \mathcal{E}_{\geq n}^{(0)}\,, 
\end{equation}
see Eq.~\eqref{semi-group}, and every $Y\in \mathcal{E}_{\geq n}^{(0)}$ can be written as
\begin{equation}\label{conjugation}
Y= \Gamma_{0}^{-n}\,X\,\Gamma_{0}^{n}, \qquad \text{  for some  }\, X\in \mathcal{E}\,,
\end{equation}
for arbitrary $n\in \mathbb{Z}_{+}$. We define 
\begin{equation}\label{commutants}
\mathcal{D}_{[n,n']}:= \big\{U(n')^{*} X U(n')\,\vert\, X \in \mathcal{D}_{[n,n']}^{(0)}\big\} \subset \mathcal{E}_{\geq n}\,, 
\quad 0\leq n<n'\,.
\end{equation}
Eqs.~\eqref{PDP-model}, \eqref{event alg}, \eqref{unit equiv} and \eqref{commutants} then imply that
$$\big[\mathcal{E}_{\geq n'}\big]' \cap \mathcal{E}_{\geq n} = \mathcal{D}_{[n,n']}\,,$$
and hence the \textit{Principle of Diminishing Potentialities} holds in the model of the interacting system.

\subsection{The law of evolution of states according to the $ETH$ Approach}
\hspace{0.5cm}\textit{``The idea that elimination of coherence, in one way or another, implies the replacement of `and' by `or', is a very common one among solvers of the `measurement problem'. It has always puzzled me.} (John Stewart Bell)\\

In this and the following subsection, it is convenient to be able to also work in the \textit{Schr\"odinger picture}, rather than 
only in the \textit{Heisenberg picture} used so far. 

\underline{Remark on the Schr\"odinger picture}: General states of $S$ are described by density matrices, 
$\widetilde{\Sigma}$, on $\mathcal{H}_S$. By
$\widetilde{\Sigma}_{n}$ we denote the density matrix obtained by restricting the state corresponding to $\widetilde{\Sigma}$ to the algebra $\mathcal{E}_{\geq n}$. By 
$\Sigma_{n}$ we denote the \textit{same} state of $S$, but viewed as a state in the \textit{Schr\"odinger picture}, which is related to the \textit{Heisenberg  picture} by the following identity: 
\begin{align}\label{Heisenberg-Schrod}
\text{tr}_{\mathcal{H}_S}\big(\widetilde{\Sigma}_{n}\cdot X(t)\big) = \text{tr}_{\mathcal{H}_S} \big(\Sigma_{n}\cdot
\big[\Gamma^{*}\big]^{t-n} \,X\, \Gamma^{t-n}\big) = \text{tr}_{\mathcal{H}_S} \big(\Gamma^{t-n}\,\Sigma_{n} \,
\big[\Gamma^{*}\big]^{t-n}\cdot X\big)\,, \quad t\geq n\,,
\end{align}
for an arbitrary operator $X \in \mathcal{E}$. 

In Eq.~\eqref{Heisenberg-Schrod},  $X(t)= \Gamma^{-t} \,X\, \Gamma^{t}$, 
$\Sigma_n := \Gamma^{n}\, \Sigma_{0}\, \Gamma^{-n}$, where $\Sigma_{0}\equiv \Sigma = \widetilde{\Sigma}\,,$  and
we recall that $\mathcal{E}_{\geq n}=\big\{\Gamma^{-n}\,X\,\Gamma^{n}\,\big|\, X\in \mathcal{E}\big\}$. \hspace{10cm}
$\square$\\

Next, we describe the \textit{time evolution of states}, as predicted by the $ETH$-Approach to QM summarized 
in Sect.~\ref{ETH}. More specifically, using Definition 6 of actual events in Subsect. 4.2 and the Collapse Postulate, {\bf{Axiom CP}}, in Subsect.~4.3, we will construct a trajectory, $\big\{\omega_n\, \big| \, n\in \mathbb{Z}_{+}\big\}$, 
of states, where $\omega_n$ is a state on the algebra $\mathcal{E}_{\geq n}$, with initial condition 
$\omega_{n=0}=\omega$, and $\omega$ is the state on the algebra $\mathcal{E}$ given by
\begin{equation}\label{initial cond}
\omega(X):= \text{tr}_{\mathcal{H}_S}\big((P_{\underline{k}} \otimes \widetilde{\Omega})\cdot X\big), \qquad X\in \mathcal{E}\,.
\end{equation}
In this equation, $P_{\underline{k}}$ denotes the orthogonal projection onto the vector 
$\Phi_{\underline{k}} \in \mathfrak{F}_S,\,  \underline{k}\in \mathcal{S}_{fin}$, and $\widetilde{\Omega}\equiv \Omega$ 
is some density matrix on the Hilbert space $\mathfrak{h}_S$ of the atom. 
We remark that when evaluated on $\mathcal{E}= \mathcal{E}_{\geq 0}^{(0)}$ (see Eq.\eqref{event algebras}) the state $\omega$ 
is \textit{independent} of $\{k_n\}_{n<0}$. 

We start our analysis by studying the restriction of the state $\omega$ defined in Eq.~\eqref{initial cond} to the algebra 
$\mathcal{E}_{\geq n}$, for some $n>0$. For this purpose, we introduce operators
($M\times M$ matrices), $\mathfrak{L}_{\alpha}^{\ell}$, on $\mathfrak{h}_S \simeq \mathbb{C}^{M}$ 
 by specifying their matrix elements
\begin{equation}\label{Kraus}
\langle u, \mathfrak{L}_{\alpha}^{\ell} v \rangle:= \langle \phi_{\alpha} \otimes u, U \,\phi_{\ell} \otimes v \rangle, \quad \alpha, \ell = 1,2,\dots, N, \,\, \,\text{ for arbitrary }\,\, u, v \text{ in  }\,\,\mathfrak{h}_S\,,
\end{equation}
where $U$ is the unitary operator chosen above Eq.~\eqref{emission op}. From expression \eqref{Kraus} for 
$\mathfrak{L}_{\alpha}^{\ell}$, the unitarity of the operator $U$ and the completeness relation 
$\sum_{\alpha=0}^{N} \vert \phi_{\alpha}\rangle\, \langle \phi_{\alpha}\vert = \mathbf{1}$ we infer that
\begin{equation}\label{sum rule}
\sum_{\alpha=1}^{N}  [\mathfrak{L}_{\alpha}^{\ell}]^{*} \cdot \mathfrak{L}_{\alpha}^{\ell} = \mathbf{1}\vert_{\mathbb{C}^{M}}\,.
\end{equation}
The operators $\big\{\mathfrak{L}_{\alpha}^{\ell}\big\}_{\alpha=1}^{N}$ are called \textit{Kraus operators} \cite{Kraus}.
Let $Z\in \mathcal{E}_{\geq n}$, then $\exists\, X\in \mathcal{E}$ such that
$$Z = \Gamma^{-n}\, X\,\Gamma^{\,n}, $$
 with  $\Gamma$  as defined in Eq.\eqref{emission op}. Any operator 
 $X\in \mathcal{E}=\mathcal{E}_{\geq 0}^{(0)}= \mathcal{A}_{\geq 0}\otimes B(\mathfrak{h}_S)$ is a linear combination 
 of operators of the form $F \otimes C$, where $F\in \mathcal{A}_{\geq 0}$ and $C$ is an $M\times M$ matrix acting on 
 $\mathfrak{h}_S$. 
 
 We first determine the restriction of the state $\omega$ to the algebra $\mathcal{E}_{\geq 1}$, (i.e., we set $n=1$). We choose an operator
 $Z:=\Gamma^{-1} (F\otimes C)\Gamma$, with $F$ and $C$ as above. The density matrix $\widetilde{\Omega}$ can be diagonalized,
 $$\widetilde{\Omega}= \sum_{j=1}^{M} p_j \vert v_j\rangle \langle v_j\vert, \qquad p_j \geq 0, \forall\, j, \quad
 \sum_{j=1}^{M}p_j =1\,,$$
 where $\big\{v_j \big\}_{j=1}^{M}$ is an orthonormal basis of eigenstates of $\widetilde{\Omega}$. 
 By inserting the partition of unity $\sum_{\alpha=1}^{N} \vert \phi_{\alpha} \rangle \langle \phi_{\alpha}\vert = \mathbf{1}\vert_{\mathbb{C}^{N}}$ and using definition \eqref{Kraus} of the operators $\mathfrak{L}_{\alpha}^{\ell}$ we show that
 \begin{align}\label{one step}
 \text{tr}_{\mathcal{H}_S}\big(&(P_{\underline{k}}\otimes \widetilde{\Omega})\cdot \Gamma^{-1}(F\otimes C)\Gamma\big) = \nonumber\\
& = \langle \Phi_{\underline{k}}, \mathfrak{S}^{-1} F\, \mathfrak{S} \Phi_{\underline{k}}\rangle \cdot \Big\{\sum_{j=1}^{M} p_{j}\,
\langle (\mathbf{1}\otimes V)U(\phi_{k_0}\otimes v_j), (\mathbf{1}\otimes C) (\mathbf{1}\otimes V) U(\phi_{k_0} \otimes v_j )\rangle\, \nonumber \\
& =\langle \Phi_{\sigma(\underline{k})}, F\,  \Phi_{\sigma(\underline{k})}\rangle \cdot \Big\{\sum_{j=1}^{M} p_{j} \sum_{\alpha=1}^{N} \langle V\mathfrak{L}_{\alpha}^{k_0} v_j, C\, V\mathfrak{L}_{\alpha}^{k_0} v_j \rangle\Big\} \nonumber \\
& = \langle \Phi_{\sigma(\underline{k})}, F\,  \Phi_{\sigma(\underline{k})}\rangle \cdot \Big\{\sum_{\alpha=1}^{N} 
\text{tr}_{\mathfrak{h}_S} \big(V\mathfrak{L}_{\alpha}^{k_0}\, \Omega\, [V\mathfrak{L}_{\alpha}^{k_0} ]^{*} \cdot C\big)\Big\}\,,
\end{align}
where $\Omega = \widetilde{\Omega}$ (at time $t=0$), and where $V$ is the $time$-1 propagator of the atom decoupled 
from the R-field.

This calculation easily generalizes to arbitrary times $n>1$. Choosing 
 $Z:= \Gamma^{-n} (F\otimes C) \Gamma^{n}$, with $F\in \mathcal{A}_{\geq 0}$ and $C\in B(\mathfrak{h}_{S})$, 
  we find that
 \begin{align}\label{formula magna}
 \omega(Z)=& \langle \Phi_{\underline{k}}, [\mathfrak{S}]^{-n} F\, \mathfrak{S}^{n} \Phi_{\underline{k}}\rangle \cdot
\Big\{ \sum_{\alpha_1, \dots, \alpha_n} \text{tr}_{\mathfrak{h}_S} \Big(\widetilde{\Omega}\, \Big\{\prod_{\ell = 1}^{n} [V\,\mathfrak{L}_{\alpha_{\ell}}^{k_{\ell}}]^{*}\Big\}\, C\, \Big\{ \prod_{j=n}^{1} (V\, \mathfrak{L}_{\alpha_j}^{k_j})\Big\} \Big) \Big\}\nonumber \\
 =& \langle \Phi_{\sigma^{n}(\underline{k})}, F \,\Phi_{\sigma^{n}(\underline{k})} \rangle\cdot \Big\{\sum_{\alpha_1, \dots, \alpha_n} 
 \text{tr}_{\mathfrak{h}_S} \Big(\Big\{V\,\mathfrak{L}_{\alpha_n}^{k_n} \cdots V\,\mathfrak{L}_{\alpha_1}^{k_1} \,\Omega\, 
 [\mathfrak{L}_{\alpha_1}^{k_1}]^{*}V^{*}\cdots [\mathfrak{L}_{\alpha_n}^{k_n}]^{*}V^{*}\Big\}\cdot C\Big) \Big\}\,,
 \end{align}
where the sum over $\alpha_j$ ranges over $\alpha_j =1,\dots, N,\,\, \forall\, j=1,\dots, n$. Formula \eqref{formula magna} 
shows that the evolution of states is entangling the state of the R-field with the state of the atom, as one 
would expect when interactions between the $R$-field and the atom are turned on. It is given by a 
\textit{quantum Markov chain}. To determine the evolution of states predicted by the $ETH$-Approach we 
will have to ``unravel'' the evolution described by formula \eqref{formula magna}; see Theorem 4 below.\\

{\bf{Lemma 3}}: \textit{The maps
\begin{equation}\label{Markov}
\Omega\,\, \mapsto V\,\Omega\,V^{*}\quad \text{ and }\quad
\Omega\,\, \mapsto \sum_{\alpha} \mathfrak{L}_{\alpha}^{\ell}\,\Omega\, [\mathfrak{L}_{\alpha}^{\ell}]^{*}
\end{equation}
are \textit{completely positive}, so that the right sides in \eqref{Markov} are again density matrices on $\mathfrak{h}_S$.}
\,$\square$\\

This lemma has been established by \textit{Kraus} in \cite{Kraus}. It implies that the map 
$$\Omega\,\,\mapsto \sum_{\alpha_1, \dots, \, \alpha_n} V\,\mathfrak{L}_{\alpha_n}^{k_n} \cdots V\,\mathfrak{L}_{\alpha_1}^{k_1} \,\Omega\, 
 [\mathfrak{L}_{\alpha_1}^{k_1}]^{*}V^{*}\cdots [\mathfrak{L}_{\alpha_n}^{k_n}]^{*}V^{*}$$
is completely positive and gives rise to a \textit{quantum Markov chain}. (We remark, in passing, that the dynamics considered in Sect.~\ref{HP} is \textit{not} Markovian, which means that it is considerably more complicated to analyze it.)\\

Next, we determine those states $\omega_n, n=0,1,2, \dots,$ on the algebras $\mathcal{E}_{\geq n}$ that can be 
reached \textit{recursively} from the initial condition $\omega_0 = \omega$, with $\omega$ as in Eq.~\eqref{initial cond}, 
by applying the \textit{law of evolution of states} specific to the $ETH$-Approach, as formulated in Definition 6 
(actual events) and {\bf{Axiom CP}} (Collapse Postulate) of Sect.~\ref{ETH}. We use induction in time to accomplish this 
task, explaining the induction step from time $m-1$ to time $m$ by outlining the construction of $\omega_m$, given 
$\omega_{m-1}$. \\
 
{\bf{Theorem 4}}: \textit{Let $Z:= \Gamma^{-n}(F\otimes C)\,\Gamma^{n} \in \mathcal{E}_{\geq n}$, with 
$F\in \mathcal{A}_{\geq 0}$ and $C\in B(\mathfrak{h}_S)$ (a general element of $\mathcal{E}_{\geq n}$ being a sum of such operators), and let $\omega_0 =\omega$ be the state on the algebra 
$\mathcal{E}$ specified in Eq.~\eqref{initial cond}. Let $\omega_n$ be a state obtained from $\omega_0$ by applying 
the law of evolution of states specific to the} $ETH$-Approach \textit{formulated in Sect.~\ref{ETH}. Then}
\begin{equation}\label{state at n}
\omega_{n}(Z)= \langle \Phi_{\sigma^{n}(\underline{k})}, F \,\Phi_{\sigma^{n}(\underline{k})}\rangle 
\cdot \text{tr}_{\mathfrak{h}_S}\big(\Omega_n \cdot C\big)\,,
\end{equation}
\textit{where} $\Omega_{n} = \big[\text{tr}_{\mathfrak{h}_S} (\Pi^{(n)})\big]^{-1} \,\Pi^{(n)}$, \textit{and $\Pi^{(n)}$ 
is an orthogonal projection on the Hilbert space $\mathfrak{h}_S$ of the atom.}

{\bf{Proof}}:
Theorem 4 is proven by induction in time $n\in \mathbb{Z}_{+}$. Eq.~\eqref{state at n} is our \textit{induction hypothesis}, denoted $(\mathcal{I}_{n})$. 
Clearly $(\mathcal{I}_{n})$ holds for $n=0$. We assume that $(\mathcal{I}_{n})$ holds for $n=m-1$, for some $m=1,2,\dots$, 
and show that this implies that it holds for $n=m$. This is done in two steps: We first restrict the state $\omega_{m-1}$ 
to the algebra $\mathcal{E}_{\geq m}\subsetneqq \mathcal{E}_{\geq m-1}$, the resulting state on $\mathcal{E}_{\geq m}$ 
being denoted by $\widehat{\omega}_m$. We then apply {\bf{Axiom CP}} (the Collapse Postulate) of Subsect.~4.3 to 
select a state $\omega_m$ subordinate to $\widehat{\omega}_m$, (i.e., $\omega_m \prec \widehat{\omega}_m$).

We now repeat steps very similar to those leading to Eq.~\eqref{one step} in more detail.
From Eq.~\eqref{unit equiv} we infer that an operator 
$\tilde{Z} \in \mathcal{E}_{\geq m}\subset \mathcal{E}_{\geq m-1}$ is a sum of operators of the form
\begin{equation}\label{shift by 1}
\tilde{Z} = U(m-1)^{*} \,\Gamma_{0}^{1-m}\, \tilde{X}\,\Gamma_{0}^{m-1} U(m-1)\,,
\end{equation}
where 
$$\tilde{X}:= U_{1}^{*}\, \big[(\mathfrak{S}^{*}F\mathfrak{S})\otimes V^{*}C\,V\big]\,U_1 \,\text{ belongs to }\, \mathcal{E}_{\geq 1}\subset \mathcal{E},$$
Thus, we can apply the induction hypothesis to $\omega_{m-1}(\tilde{Z})$. 
The density matrix $\Omega_{m-1}$ can be written as
\begin{equation}\label{matrix}
\Omega_{m-1}= \sum_{j=1}^{M} p_j \vert v_j^{(m-1)}\rangle \langle v_j^{(m-1)} \vert\,,
\end{equation}
where $\big\{ v_{j}^{(m-1)} \big\}_{j=1}^{M}$ is a complete orthonormal system of eigenstates of $\Omega_{m-1}$, $p_j \geq 0,\,\forall \,j=1,\dots, M,$ with \mbox{$\sum_{j=1}^{M}p_j =1$.}
Note that the induction hypothesis $(\mathcal{I}_{m-1})$ is linear in $\Omega_{m-1}$; see Eq.~\eqref{state at n}. Thus, for $Z$ as in the statement of Theorem 4, $(\mathcal{I}_{m-1})$ can be written as
\begin{align}\label{expand}
\omega_{m-1}(Z)&=\sum_{j=1}^{M} p_j \,\omega_{m-1}^{j}(Z)\,, \qquad \text{where} \nonumber \\
 \omega_{m-1}^{j}(Z)&:= \langle \Phi_{\sigma^{m-1}(\underline{k})}, F\,\, \Phi_{\sigma^{m-1}(\underline{k})} \rangle \cdot \langle v_{j}^{(m-1)}, C\,v_{j}^{(m-1)}\rangle\,. 
\end{align}
If we now set $Z:=\tilde{Z}$, with $\tilde{Z}$ as specified in \eqref{shift by 1}, the $j^{th}$ term on the right side is given by
\begin{equation}\label{jth term}
\omega_{m-1}^{j}(\tilde{Z})=\langle \Phi_{\sigma^{m}(\underline{k})}, F\,\, \Phi_{\sigma^{m}(\underline{k})} \rangle \cdot
\langle U\,[ \phi_{\sigma^{m-1}(\underline{k})_{1}}\otimes \,v_{j}^{(m-1)} ], \big(\mathbf{1}\otimes 
V^{*}\,C\cdot V \big)\,U\,[\phi_{\sigma^{m-1}(\underline{k})_{1}}\otimes \,v_{j}^{(m-1)} ]\rangle
\end{equation}
We recall that $\sigma^{m-1}(\underline{k})_1 = k_{m-1}$. Next, we recall the definition of the Kraus operators 
$\mathfrak{L}_{\alpha}^{\ell}$ (see Eq.~\eqref{Kraus}) and use the completeness of 
$\big\{\phi_{\alpha}\big\}_{\alpha=1}^{N}$ to show that
$$\omega_{m-1}^{j}(\tilde{Z})= \langle \Phi_{\sigma^{m}(\underline{k})}, F\,\, 
\Phi_{\sigma^{m}(\underline{k})} \rangle \cdot \Big\{ \sum_{\alpha=1}^{N} \langle V\,\mathfrak{L}_{\alpha}^{{k}_{m-1}}\,
v_{j}^{(m-1)}, C\,V\,\mathfrak{L}_{\alpha}^{{k}_{m-1}}\, v_{j}^{(m-1)}\rangle\Big\}\,.$$
Defining
\begin{equation}\label{matrix-n}
\widehat{\Omega}_{m}:= \sum_{\alpha=1}^{N} V\cdot \mathfrak{L}_{\alpha}^{k_{m-1}} \, \Omega_{m-1}
\big(\mathfrak{L}_{\alpha}^{k_{m-1}}\big)^{*}\cdot V^{*}\,,
\end{equation}
and recalling \eqref{matrix} we conclude that
\begin{equation}\label{time evol}
\widehat{\omega}_{m}(\tilde{Z})\equiv \omega_{m-1}(\tilde{Z})= \langle \Phi_{\sigma^{m}(\underline{k})}, F\,\, \Phi_{\sigma^{m}(\underline{k})} \rangle \cdot 
\text{tr}_{\mathfrak{h}_S} \big(\widehat{\Omega}_{m} C\big)\,.
\end{equation}
Lemma 3 tells us that the right side of \eqref{matrix-n} defines a density matrix on $\mathfrak{h}_S$. Let
\begin{equation}\label{spect dec}
\widehat{\Omega}_{m} = \sum_{r=1}^{s} q_{r}^{(m)} \Pi_{r}^{(m)}, \qquad s\leq M\,,
\end{equation}
be the spectral decomposition of $\widehat{\Omega}_m$, with $\Pi_{r}^{(m)}$ the spectral projection of 
$\widehat{\Omega}_m$ corresponding to the eigenvalue $q_{r}^{(m)}$. We order the eigenvalues such that
$$q_{1}^{(m)} > q_{2}^{(m)} > \dots q_{s}^{(m)} > 0,\quad  \text{and we notice that }\quad
\sum_{r=1}^{s} q_{r}^{(m)} \,\text{tr}_{\mathfrak{h}_S}\big(\Pi_{r}^{(m)}\big) =1.$$
Let $P_{\underline{k}}$ be the rank-1 orthogonal projection onto $\Phi_{\underline{k}}$ and 
$P_{\underline{k}}^{\perp}: =\mathbf{1}- P_{\underline{k}}$. Using Definition 6 (actual events) of Sect.~\ref{ETH}, 
we find that, in the \textit{Schr\"odinger picture}, the actual event happening at time $m$ is described  
by the family of orthogonal projections 
\begin{equation}\label{event m}
\big\{P_{\sigma^{m}(\underline{k})} \Pi_{r}^{(m)}, P_{\sigma^{m}(\underline{k})}^{\perp} \Pi_{r}^{(m)}\big\}_{r=1}^{s}\,,
\end{equation}
which generate an algebra unitarily conjugated to the center, $\mathcal{Z}_{\widehat{\omega}_m}(\mathcal{E}_{\geq m})$, 
of the centralizer of $\widehat{\omega}_m$, with $\widehat{\omega}_m :=\omega_{m-1}\vert_{\mathcal{E}_{\geq m}}$. We now apply {\bf{Axiom CP}} (the Collapse Postulate) formulated in Subsect.~4.3: 
The probability of the state $\widehat{\omega}_m$ collapsing onto the range of a projection proportional to 
$P_{\sigma^{n}(\underline{k})}^{\perp}$ vanishes, as follows from \eqref{time evol} and \textit{Born's Rule} 
(see Eq.~\eqref{Born Rule}, Subsect.~4.3). We thus conclude that, when the event described in 
Eq.~\eqref{event m} sets in at time $m$, the state of the system collapses onto one of the states
\begin{align}\label{state m}
\omega_{m}^{(r)}(\tilde{Z})=&  \langle \Phi_{\sigma^{m}(\underline{k})}, F\,\, \Phi_{\sigma^{m}(\underline{k})} \rangle \cdot 
\text{tr}_{\mathfrak{h}_S}\big(\Omega_{m}^{(r)} \cdot C\big)\,,\,\,\text{ where }\nonumber \\
\Omega_{m}^{(r)}:=&\big[\text{tr}_{\mathfrak{h}_S}\big(\Pi_{r}^{(m)}\big)\big]^{-1}\, \Pi_{r}^{(m)}\,, \,\,\, \text{ for some }\,\, 
r \in \big\{1, \dots, s\big\} .
\end{align}
According to {\bf{Axiom CP}}, the probability to choose the state $\propto \Pi_{r}^{(m)}$ is given by
$$\text{prob}\{r\}= q_{r}^{(m)} \cdot \text{tr}_{\mathfrak{h}_S}\big(\Pi_{r}^{(m)}\big)\qquad (Born's Rule)$$
Eq.~\eqref{state m} implies that, in the \textit{Schr\"odinger picture}, where ``observables'' are taken to 
be \textit{time-independent}, the state, $\Omega_m$, of the atom at time $m$ is given by one of the states 
$\Omega_{m}^{(r)}$. 

Eqs.~\eqref{matrix-n} and {\eqref{state m} complete the induction step, $(\mathcal{I}_{m-1}) \Rightarrow (\mathcal{I}_m)$.\hspace{3.6cm}$\square$\\

We note that, when restricted to operators that are functions of the $R$-field, but act trivially on the Hilbert space 
$\mathfrak{h}_S$ of the atom, the states $\omega_{n},n=0,1,\dots,$ are \textit{product states}. This implies that the effective time evolution of the state of the atom described above is \textit{``Markovian''}. Moreover, if the atom is \textit{decoupled} from the $R$-field, corresponding to $U=\mathbf{1}$ in Eq.~\eqref{emission op}, then 
$$\Omega_n = V^{n}\Omega\, V^{-n}, \qquad \text{with }\,\, \Omega=\widetilde{\Omega} \,\,\text{ as in }\,\,\eqref{initial cond},$$
i.e., the evolution of states of the atom is governed by \textit{Sch\"odinger-Liouville evolution} – the atom decoupled from the $R$-field is a perfectly closed system.
\subsection{A more concrete model of an atom interacting with the $R$-field}\label{concrete model}
\hspace{0.5cm}\textit{``The concepts `system', `apparatus', `environment', immediately imply an artificial division of 
the world, and an intention to neglect, or take only schematic account of, the interaction across the split.''} (John Stewart Bell)

It is instructive to study an example of an explicit operator $U$ describing interactions between the atom and the $R$-field 
(see Eq.~\eqref{emission op}): We choose a partition of unity, $\big\{Q_m\big\}_{m=1}^{L}\,,$ with\, $L\leq M$, by orthogonal 
projections acting on $\mathfrak{h}_S\simeq \mathbb{C}^{M}$ and define $U$ by setting
\begin{equation}\label{measurement op}
U:= \sum_{m=1}^{L} T^{(m)}\otimes Q_{m}, 
\end{equation}
where $T^{(m)}$ is a unitary operator on $\mathbb{C}^{N}$, while \, $Q_m=Q_m^{*}$ is an orthogonal projection on 
$\mathfrak{h}_S \simeq \mathbb{C}^{M}$\,, \,\,$Q_m \cdot Q_{\ell} = \delta_{m \ell}\, Q_m, \,\forall \,m, \ell =1,\dots, M,
 \,\, \text{ and }\,\,\sum_{m=1}^{L} Q_m = \mathbf{1}\,.$ For this choice of $U$ we find that
\begin{equation}\label{Kraus example}
\mathfrak{L}_{\alpha}^{k_{\ell}}= \sum_{m=1}^{L} \langle \phi_{\alpha}, T^{(m)} \phi_{k_{\ell}}\rangle Q_{m}\,.
\end{equation}
Let $\Omega_{n-1}$ be the density matrix describing the state of the atom at time $n-1$. Then the state 
$\widehat{\Omega}_n$ of the atom at time $n$, obtained by restricting $\omega_{n-1}$ to the algebra 
$\mathcal{E}_{\geq n}$, is given by Eq.~\eqref{matrix-n} (with $m\rightarrow n$), namely
\begin{equation}\label{step-1-evol}
\widehat{\Omega}_n = \sum_{\ell, m =1, \dots, L} g^{\ell m}(n-1) \, V\,Q_{\ell}\, \Omega_{n-1}\, Q_mV^{*}\,,
\end{equation}
where
\begin{equation}\label{metric}
g^{\ell m}(j):= \langle \phi_{k_j}, (T^{(m)})^{*} \,T^{(\ell)} \phi_{k_j}\rangle = \langle \,T^{(m)}\phi_{k_j}, T^{(\ell)} \phi_{k_j}\rangle\,. \end{equation}
This is a direct consequence of Eqs.~\eqref{matrix-n}, \eqref{Kraus example} and the completeness of $\big\{\phi_{\alpha}\big\}_{\alpha=1}^{N}$.

We note that, for an arbitrary time $j$,
\begin{align}\label{hermiticity}
\overline{g^{\ell m}(j)}= \langle\, T^{(\ell)} \phi_{k_j} , T^{(m)} \phi_{k_j} \rangle& = g^{m \ell} (j)\,,\quad\quad g^{mm}(j) = 1,\,\,\forall\,m=1,\dots, L,\,\, \text{ and }\nonumber \\
&\sum_{\ell, m=1}^{L} \overline{v}_{\ell}\, g^{\ell m}(j) \,v_{m} \geq 0\,,
\end{align}
for an arbitrary $L$-tuple, $v:=(v_1,\dots, v_{L})$, of complex numbers; i.e., the matrix 
$$G(j) := \Big(g^{\ell m}(j)\Big)_{\ell, m= 1}^{L}$$ 
is a (hermitian) non-negative matrix on $\mathbb{C}^{L}$ and hence can be diagonalized by a unitary $L\times L$ matrix, 
$D(j)=\big(d_{r}^{\,s}(j)\big)_{r,s=1}^{L} $:
\begin{align}\label{diag}
\hspace{1cm}\gamma_{r}(j)\, \delta^{rs}= \sum_{\ell m}\, &\overline{d_{\ell}^{\, r}(j)} \, g^{\ell m}(j) \,d_{m}^{\, s}(j)\,,\nonumber \\
i.e., \hspace{0.6cm} \text{diag}\Big(\gamma(j)&\Big) = D(j)^{*}\,G(j)\, D(j)\,. \hspace{2.4cm}
\end{align}
The non-negative numbers $\gamma_{r}(j)$ are the eigenvalues of the matrix $G(j)$, and we have that 
$$\sum_{r=1}^{L} \gamma_{r}(j) = \text{tr}\big(G(j)\big) = \sum_{m=1}^{L}g^{mm}(j) = L\,.$$
Notice that if $(M\geq) L > N$ then 
$L-N$ eigenvalues of $G(j)$ necessarily vanish. (This is because the vectors $\big\{T^{(\ell)} \phi_{k_j} \,\big|\, \ell=1, \dots L\big\}$ are necessarily linearly dependent if $L>N$.) Physically, it is, however, more realistic to suppose that $N\gg M$. 
We will see that if one of the eigenvalues $\big\{\gamma_{r}(n-1)\big\}_{r=1}^{L}$ is very close to $L$ then the map 
$\Omega_{n-1} \mapsto \widehat{\Omega}_n$ is close to being given by conjugation with a unitary matrix.

Eq.~\eqref{step-1-evol} can be cast into the following form: Let $\Omega_{n-1}$ denote the density matrix describing the state of the atom at time $n-1$. Then the density matrix describing 
the state of the atom at time $n$, \textit{before {\bf{Axiom CP}} is applied}, is given by
\begin{equation}\label{state rec}
\widehat{\Omega}_{n}= \sum_{r=1}^{L} \gamma_{r}(n-1)\,V \mathfrak{K}_{r}(n-1)\,\Omega_{n-1}\, \mathfrak{K}_{r}(n-1)^{\,*}\,V^{*}\,,\,\,\, \text{ where }\,\mathfrak{K}_{r}(n-1):= \sum_{m=1}^{L} d_{r}^{\,m}(n-1)Q_{m}\,.
\end{equation}
This equation shows that $\widehat{\Omega}_n$ is non-negative, and using that 
$g^{mm}(n-1) =1,\,\forall\, m=1, \dots, L$, we see that it has trace equal to $1$. Thus, $\widehat{\Omega}_{n}$ is again 
a density matrix, which can be written as a convex combination of disjoint orthogonal projections, $\Pi_{r}^{(n)}$, as in 
Eq.~\eqref{spect dec} (with $m\rightarrow n$). Applying {\bf{Axiom CP}}, we recover an expression equivalent to the 
one in Eq.~\eqref{state m}.\\

{\bf{The weak-coupling regime of the model}}\\

It is interesting to study some limiting regimes in the model introduced above. We first consider the \textit{weak-coupling 
regime}, which is characterized by
\begin{equation}\label{weak coupling}
T^{(m)}= \mathbf{1} + \varepsilon\cdot \tau^{(m)}, \qquad \text{ with }\,\, \Vert \tau^{m}\Vert\,\leq 1,\,\,\forall\, m=1,\dots, L\,,\,\,
 \text{and }\,\,0\leq \varepsilon \ll 1.
 \end{equation}
It is easy to see that this implies that, for arbitrary $j$,
\begin{equation}\label{size of ev}
g^{\ell m}(j) = 1 + \mathcal{O}(\varepsilon),\qquad \forall\,\, \ell, m\,, 
\end{equation}
and
$$\gamma_{1}(j) = L+\mathcal{O}(\varepsilon)\,, \,\,\, \gamma_{r}(j) =\mathcal{O}(\varepsilon)\,, \,\,\forall\, r>1\,,\,\,\,
d_{1}^{\,s}(n) = \frac{1}{\sqrt{L}} \big(1+ \mathcal{O}(\varepsilon)\big)\,, \forall\, s\,.$$
Eq.~\eqref{step-1-evol}, combined with \,$\sum_{m=1}^{L} Q_m =1$,  then implies that
\begin{equation}\label{standard}
\Vert \widehat{\Omega}_{n}- V\, \Omega_{n-1}\, V^{*}\Vert = \mathcal{O}(\varepsilon)\,, \quad i.e.,\, \, \,\,
\widehat{\Omega}_n \approx V\,\Omega_{n-1}\,V^{*}\,.
\end{equation}
According to the Collapse Postulate, 
$$\Omega_{n-1}= \mathcal{N}^{-1} \Pi, \text{ where }\, \Pi=\Pi^{*}= \Pi^{2}\,, \quad \mathcal{N}= \text{tr}_{\mathfrak{h}_S}(\Pi)\,,$$
i.e., $\Pi$ is an orthogonal projection. Eqs.~\eqref{step-1-evol} and \eqref{size of ev} then imply that
\begin{align}\label{approx proj}
\widehat{\Omega}_{n}=&q^{(n)} \Pi^{(n)} + \sum_{r\geq 2} q_{r}^{(n)} \Pi_{r}^{(n)}, \qquad \text{where  }\nonumber\\
q^{(n)}\equiv q_{1}^{(n)} = \mathcal{N}^{-1}+ \mathcal{O}(\varepsilon), &\quad q_{r}^{(n)} = \mathcal{O}(\varepsilon),\,\,r\geq 2, \,\, \text{ and } \,\, 
\Vert  \Pi^{(n)} -  V\,\Pi V^{*} \Vert = \mathcal{O}(\varepsilon)\,.
\end{align}
The Collapse Postulate ({\bf{Axiom CP}} of Sect.~\ref{ETH}) implies that, with \textit{very high probability}
$$\Omega_{n} = \big[\text{tr}_{\mathfrak{h}_S}\big(\Pi^{(n)}\big)\big]^{-1} \cdot \Pi^{(n)} \simeq V\, \Omega_{n-1} V^{*}\,.$$
Thus, the system obtained by tracing out the $R$-field is well approximated by the closed system consisting of just the atom 
(decoupled from the $R$-field), whose states evolve unitarily by conjugation with powers of the operator $V$. However,
every once in a while, it will happen -- for \textit{purely entropic reasons} -- that the state of the system collapses to a very 
unlikely state $\Omega_{n} \propto \Pi^{(n)}_{r}$, for some $r\geq 2$, with $\Pi_{r}^{(n)}$ approximately orthogonal to 
$V\,\Pi\, V^{*}$, which represents a strong deviation from unitary 
evolution. An observer will perceive a collapse to such an unlikely state as an {\bf{event}} in the literal sense of the word. 
The frequency of collapse to an unlikely state is proportional to $\varepsilon$.\\

{\bf{The strong-coupling regime of the model}}\\

The strong coupling limit is characterized by the property that
\begin{align}\label{strong coupling}
g_{k}^ {\ell m} := \langle T^{m} \phi_{k}, T^{\ell} \phi_{k} \rangle = \delta^{\,\ell m} + \mathcal{O}(\varepsilon)\,, \quad \text{with }\,\,\varepsilon \ll 1\,,
\end{align}
for some or all of the vectors $\phi_{k}$, in particular for $\phi_0$. Given a state vector, 
$\Phi_{\underline{k}}\in \mathfrak{F}_S$, of the $R$-field, we set $g^{\ell m}(j)= g_{k_j}^{\ell m}$, 
as in Eq.~\eqref{metric}. Since, for our choice of a reference vector, $\Phi_{\underline{0}}$, in the construction 
of the Hilbert space $\mathfrak{F}_S$, we have that $k_j = 0$, except for finitely many values of $j$, 
the following considerations apply to the analysis of evolution of states at large times under the only assumption that \eqref{strong coupling} holds for $k =0$. 
It follows from Eq.~\eqref{step-1-evol} that if \eqref{strong coupling} holds for $k=k_{n-1}$ then
\begin{equation}\label{Markovian}
\widehat{\Omega}_n = \sum_{m=1}^{L} V\,Q_m \Omega_{n-1} Q_m \,V^{*}+ \mathfrak{e}_{n}(\varepsilon)\,, \quad \text{with }
\,\, \Vert \mathfrak{e}_{n}(\varepsilon) \Vert = \mathcal{O}(\varepsilon),
\end{equation}
where $\mathfrak{e}_n(\varepsilon)$ is some hermitian $M\times M$ matrix. It then follows from {\bf{Axiom CP}} 
of Sect.~\ref{ETH}  that $\Omega_n$ is proportional to a spectral projection of $\widehat{\Omega}_n$.
We note that if $n$ is large enough (depending on the sequence $\underline{k}$) then $k_j =0, \,\forall \, j\geq n-1$, 
and hence $g^{\ell m}(j) = g_{k_j}^{\ell m} = g_{0}^{\ell m}$ satisfies \eqref{strong coupling}, for all $j \geq n-1$. 

\underline{Remark}: The map
\begin{equation}\label{CP}
\Omega\, \mapsto\, \widehat{\Omega} := \sum_{m =1}^{L} V\,Q_{m}\, \Omega \, Q_{m} \, V^{*}, 
\end{equation}
is completely positive; (the operators $\big\{\mathfrak{L}_{m}:=V\,Q_{m}\,\vert\, m=1,\dots, L\big\}$ are Kraus operators). \\

Next, we consider the following special choice of a partition of unity $\big\{Q_{m}\big\}_{m=1}^{L}$:
\begin{equation}\label{rank 1}
Q_m = \vert \psi_m \rangle \langle \psi_m \vert\,, \qquad m=1,\dots, L, \,\text{ with }\, L=M=\text{dim}\mathfrak{h}_S\,,
\end{equation}
where $\big\{\psi_m\big\}_{m=1}^{M}$ is an orthonormal basis of $\mathfrak{h}_S$. We define a 
\textit{transition matrix (or -function)}, 
$P=\big(P(\ell, m)\big)_{\ell, m=1}^{M}$, by setting
\begin{equation}\label{trans funct}
P(\ell, m):= \vert \langle \psi_\ell, V \psi_m \rangle \vert^{2} \geq 0, \qquad \ell, m = 1,\dots, M\,.
\end{equation}
The completeness of the vectors $\big\{\psi_m \vert \,m=1,\dots, M\big\}$ and the unitarity of $V$ imply that
\begin{equation}\label{proba}
\sum_{m=1}^{M} P(\ell,m) = \sum_{\ell=1}^{M} P(\ell, m) =1\,.
\end{equation}
Using \eqref{Markovian} and \eqref{CP}, we find that, in the strong-coupling regime and for sufficiently large times,  
the time evolution of the state of the atom in the Schr\"odinger picture is well approximated by a trajectory of states 
$\big\{ \psi_{\xi_{n}} \,\vert\, 1\ll n\in \mathbb{Z}_{+}\big\}$ indexed by a {\bf{sample path}}, 
$\big\{\xi_{n} \in \mathfrak{X}\,\vert\, n \in \mathbb{Z}_{+} \big\}$, of the Markov chain with state space 
$\mathfrak{X}:=\big\{1, \dots, M\big\}$ and transition matrix $P$ defined in \eqref{trans funct}. The probabilities
$$\mu(m) := \text{prob}\big\{\text{the atom occupies state } \psi_m\big\}, \text{ for }\,m=1,\dots, M,$$
on the state space $\mathfrak{X}$ of the Markov chain evolve approximately according to 
\begin{equation}\label{Markov chain}
\mu_{n}(\ell)= \sum_{m=1}^{M} P(\ell, m)\, \mu_{n-1}(m)\,, \qquad \mu_{n-1}(m)\geq 0, \, \forall\, m, \quad \sum_{m=1}^{M}\mu_{n-1}(m)=1\,,
\end{equation}
as can be inferred from Eq.~\eqref{CP}. The positive number $P(\ell, m)$ can be interpreted as the approximate value 
of the probability of the event that the atom occupies state $\psi_{\ell}$ at some time $n$, assuming that at time $n-1$ 
it has occupied state $\psi_{m}$.

Recalling that, in the weak-coupling regime, the Schr\"odinger-picture time evolution of states of the 
atom is well approximated by unitary evolution -- the one we are used to from text books on elementary Quantum 
Mechanics -- we find that, in the models considered here, the law of evolution of states of the atom in the $ETH$-Approach 
smoothly interpolates between unitary deterministic Schr\"odinger evolution, appropriate for closed systems, and classical 
Markovian evolution of the (state-occupation) probabilities $\mu(\cdot)$, appropriate for isolated open systems of matter very strongly coupled to the radiation field.\\

{\bf{Alternation between unitary evolution and state collapse in measurements}}\\

To conclude this subsection, we briefly sketch how, in suitable situations, the alternation between linear unitary 
Schr\"odinger evolution of states of a system and non-linear state collapse
 in \textit{measurements}, as stipulated in the Copennhagen Interpretation of QM, can be understood as 
 an \textit{approximation} to the fundamental law of evolution of states in the $ETH$-Approach.

We consider models of the kind introduced in Eqs.~\eqref{measurement op} - \eqref{metric}. 
 Let $U:= \sum_{m=1}^{L} T^{(m)}\,Q_{m}$\,, see  Eqs.~\eqref{emission op} and \eqref{measurement op}.
We decompose the Hilbert space $\mathfrak{h}_S$ of the atom into a direct sum
\begin{equation}\label{dec}
\mathfrak{h}_S = \mathfrak{h}^{w} \oplus \mathfrak{h}^{s}, 
\end{equation}
with dim($\mathfrak{h}^{w}) = K < M$, and we assume that the ranges of the projections $Q_1, \dots, Q_J, J\leq K,$ 
\mbox{are contained } in $\mathfrak{h}^{w}$, while the ranges of $Q_{J+1}, \dots Q_L$ are contained in $\mathfrak{h}^{s}$. 
We interpret the numbers
$$g^{\ell m}(j):=  \langle \,T^{(m)}\phi_{k_j}, T^{(\ell)} \phi_{k_j}\rangle$$
as the matrix elements of an $L\times L$ matrix, $G(j)$ (see below \eqref{hermiticity}), acting on the vector space 
$$\mathcal{V}:=\mathbb{C}^{L}= \mathcal{V}^{w} \oplus \mathcal{V}^{s}\,,\,\, \text{ where }\,\, \mathcal{V}^{w}\simeq 
\mathbb{C}^{J},\,\, \text{ and }\,\, \mathcal{V}^{s}\simeq \mathbb{C}^{L-J}\,,$$
$\mathcal{V} \ni \mathbf{v}=(v_1,\dots, v_L) = (\mathbf{v}^{w}, \mathbf{v}^{s})$, with 
$\mathbf{v}^{w}=(v_1, \dots, v_J)\in \mathcal{V}^{w}$ and $\mathbf{v}^{s}=(v_{J+1}, \dots, v_{L})\in \mathcal{V}^{s}$.
The matrix $G(j)$ is assumed to have the property that
\begin{align}\label{metric ex}
G(j) = G_{0} + \Delta G(j), \qquad G_{0}= G_{0}^{w}\vert_{\mathcal{V}^{w}} \oplus G_{0}^{s}\vert_{\mathcal{V}^{s}}\,,
\end{align}
where
\begin{equation}\label{matrices}
G_{0}^{w} = \begin{pmatrix} 1& \hdots & 1\\ \vdots & & \vdots \\ 1& \hdots &1 \end{pmatrix}\,, \quad
G_{0}^{s} = \mathbf{1}\vert_{\mathcal{V}^{s}}\quad \text{  and  }\quad \Vert \Delta G(j) \Vert \leq \varepsilon \ll 1\,.
\end{equation}
We also assume that the \textit{time-1} propagator $V$ of the atom has the property that 
\begin{equation}\label{slow evol}
V = V_{0} + \Delta V, \quad \text{ where }\,\, V_{0}= V_{0}\vert_{\mathfrak{h}^{w}} \oplus \mathbf{1}\vert_{\mathfrak{h}^{s}}\,, \,\,\,\text{and }\, \,\,\Vert \Delta V \Vert \leq \delta\,,
\end{equation}
for some $\delta \ll 1$, i.e., it takes a long time of $\mathcal{O}(\delta^{-1})$ for a state prepared in the subspace 
$\mathfrak{h}^{w}$ to develop a substantial overlap with a state in the subspace $\mathfrak{h}^{s}$, and the 
propagator of the atom restricted to the subspace $\mathfrak{h}^{s}$ is very close to the identity operator. 

Let us suppose that the initial state of the atom is given by $\Omega_{0}:=\big[\text{tr}(\Pi)\big]^{-1} \Pi$, 
where $\Pi$ is an orthogonal projection whose range is contained in $\mathfrak{h}^{w}$, i.e., $\Pi\vert_{\mathfrak{h}^{s}} = 0$, 
meaning that the initial state of the atom belongs to the subspace $\mathfrak{h}^{w}$ of states only 
very \textit{weakly} coupled to the $R$-field. Eq.~\eqref{slow evol} then implies that the state of the 
atom will remain in the subspace $\mathfrak{h}^{w}$ for a period of time of duration $\mathcal{O}(\delta^{-1})$, with 
only tiny tails leaking into the subspace $\mathfrak{h}^{s}$. The form of the matrix $G_{0}^{w}$ given in \eqref{matrices} 
and the fact that $\Vert \Delta G(j) \Vert \leq \varepsilon \ll 1$ then imply that the evolution of the state of the atom 
with initial condition $\Omega_{0}$ is well approximated by unitary Schr\"odinger evolution, as determined by the
\textit{time-1} propagator $V=V_0 +\mathcal{O}(\delta)$ of the atom, for a length of time of $\mathcal{O}(\delta^{-1})$, until the state of the atom  
develops a substantial overlap with the subspace $\mathfrak{h}^{s}$. {\bf{Axiom CP}} of Sect.~\ref{ETH} tells us that, 
once the state of the atom has a substantial overlap with $\mathfrak{h}^{s}$, it becomes likely that it collapses 
onto a state, $\Omega^{s}$, with only a tiny overlap with the subspace $\mathfrak{h}^{w}$. Assumption \eqref{matrices} 
then implies that the \textit{strong-coupling} law in Eq.~\eqref{Markovian} governs the further evolution of the state 
of the atom for a period of time of $\mathcal{O}(\delta^{-1})$. Assumption \eqref{slow evol} then entails that the state
of the atom collapses to a projection in the range of one of the projections $Q_m$, with $J+1 \leq m\leq L$, and stays
there for a period of time of duration $\mathcal{O}(\delta^{-1})$. This can be interpreted as a {\bf{measurement}} taking place, 
with a ``measurment basis'' consisting of the ranges of the projections $Q_{J+1}, \dots, Q_L$. 

Our discussion shows that the \textit{time} when the state of the atom collapses from a density matrix whose range belongs 
to the subspace $\mathfrak{h}^{w}$ of states weakly coupled to the $R$-field to one whose range belongs to the subspace 
$\mathfrak{h}^{s}$ of states strongly coupled to the $R$-field, signaling the onset of a measurement, is a 
\textit{random variable}, i.e., it is \textit{not} determined sharply by the theory. Its \textit{distribution/law} is, however, 
predicted by the theory. In other words, the question \textit{``when does the detector click?''} is answered by saying 
that the time when it clicks is a random variable whose distribution can however be determined.

 The ideas described here can be incorporated into full-fledged models of measurements performed on micro-systems, such as atoms or molecules, which  are only very \textit{weakly} coupled to the $R$-field, but will, through interactions, get entangled with \textit{measuring instruments}, the latter being quantum-mechanical systems \textit{strongly} coupled to the $R$-field (except when they are in their ``ground-state''). Explicit examples of models of measurements and measurement instruments will be communicated in a separate paper. \\
 
 {\bf{Other choices of reference states for the $R$-field}}\\
 
To conclude this section we comment on other possible choices of reference vectors, in particular thermal states, in the construction of the Hilbert space of states of the $R$-field. 

 An arbitrary operator in the algebra $\mathcal{A}_{[n,n']}, 0\leq n<n',$ is given by a sum of products of operators $A_j$ acting as the identity on all spaces $\mathcal{H}_{\ell}, \ell \not= j,$ and as an $N\times N$  matrix, also denoted by $A_j$, on the space $\mathcal{H}_{j} \simeq \mathbb{C}^{N}$, for $j=n, \dots, n'-1$. An algebra $\mathcal{A}$ is defined by
$$\mathcal{A}:= \bigvee_{n<n'<\infty} \mathcal{A}_{[n,n']}\,.$$
We choose a density matrix, $\Phi$, on $\mathbb{C}^{N}$ by setting
\begin{equation}\label{density matrix}
\Phi = \sum_{k=1}^{K} p_{k} \,\big[\text{tr}(P_{k})\big]^{-1} P_{k} \,, \quad 1\geq p_1 > \dots > p_K > 0\,, \quad 
\sum_{k=1}^{K} p_{k} =1\,, \,\,\, K\leq N\,,
\end{equation}
where $\big\{P_{k}\big\}_{k=1}^{K}$ is a family of orthogonal projections on $\mathbb{C}^{N}$, with 
$P_{k}\cdot P_{\ell}=\delta_{k \ell} P_{k}\,,\, \forall\, k, \ell$, and 
$\sum_{k} P_{k} \leq \mathbf{1}\vert_{\mathbb{C}^{N}}$. For later purposes, we set 
$P_0 := \mathbf{1}\vert_{\mathbb{C}^{N}}$. Physically, the density matrix $\Phi$ may describe a thermal 
state of the $R$-field in a time slice.

We define a product state $\varphi$ on $\mathcal{A}$ by setting
\begin{equation}\label{mixture}
\varphi(A)= \prod_{j=n}^{n'} \text{tr}\big(\Phi\cdot A_j\big), \quad \text{ for }\,\, \,A=\bigotimes_{j=n}^{n'}A_j \in \mathcal{A}_{[n,n']}\subset \mathcal{A}\,, \quad\,\, n<n'< \infty \,,
\end{equation}
with $\Phi$ as in \eqref{density matrix}.

A Hilbert space, $\mathfrak{F}_{\varphi}$, of state vectors of the $R$-field is obtained by applying the so-called 
\textit{GNS construction} to the pair $\big(\mathcal{A}, \varphi \big)$ (see, e.g., \cite{Takesaki}). 
The space $\mathfrak{F}_{\varphi}$ carries a $^{*}$representation, $\pi_{\varphi}$, of $\mathcal{A}$; in the following, we will 
not distinguish between $A$ and $\pi_{\varphi}(A)$, for $A\in \mathcal{A}$. 
For an operator $A=\bigotimes_{j=n}^{n'}A_j,\, \text{ with }\,\,A_j \in B(\mathcal{H}_j),$ we define
\begin{equation}\label{automorphism}
\sigma(A):= \bigotimes_{j=n}^{n'} A_{j}\vert_{\mathcal{H}_{j+1}}\,.
\end{equation}
This defines a $^{*}$automorphism of the algebra $\mathcal{A}$. It is obvious that the state $\varphi$ is invariant under 
$\sigma$, i.e.,
$$\varphi\big(\sigma(A)\big)=\varphi(A), \quad \forall A\in \mathcal{A}\,,$$
which implies  that there is a unitary operator $\mathfrak{S}$ acting on $\mathfrak{F}_{\varphi}$ such that
\begin{equation}\label{implementation}
\sigma(A)= \mathfrak{S}^{-1} \,A\,\mathfrak{S}\,, \quad \forall \, A\in \mathcal{A}\,.
\end{equation}
The operator $\mathfrak{S}$ generates a unitary propagator on $\mathfrak{F}_{\varphi}$ for the $R$-field in the Schr\"odinger picture.

We introduce event algebras
$$\mathcal{A}_{\geq n}:= \overline{\bigvee_{n' >n} \pi_{\varphi}\big(\mathcal{A}_{[n,n']}}\big)\,,$$
where the closure is taken in the topology of weak convergence of operators on $\mathfrak{F}_{\varphi}$.

As before, we monitor the evolution of the system only for times $n\geq 0$. In order to find the law of time evolution of 
states conforming to the $ETH$-Approach (see Definition 6 and {\bf{Axiom CP}}, Sect.~\ref{ETH}), we have to 
determine the center, $\mathcal{Z}_{\varphi}(\mathcal{A}_{\geq n})$, of the centralizer of the state $\varphi$ 
restricted to the algebra $\mathcal{A}_{\geq n}$, for an arbitrary $n\geq 0$. Since $\varphi$ is a time-translation 
invariant product state, the value of $n$ is unimportant. Among orthogonal projections belonging to 
$\mathcal{Z}_{\varphi}(\mathcal{A}_{\geq n})$ are all the operators
\begin{equation}\label{central proj}
\pi_{\underline{k}} := \bigotimes_{j\geq n} P_{k_j}\vert_{\mathcal{H}_j}, \quad \text{  with  }\,\,\underline{k}\in \mathcal{S}_{fin}\,,
\end{equation}
where $\mathcal{S}_{fin}$ is the set of sequences $\underline{k}=\big\{k_j\big\}_{j=0}^{\infty}$ with the property that 
$k_j =0$, except for finitely many $j\in \mathbb{Z}_{+}$.
It is then not difficult to see that the spectrum, $\mathfrak{X}$, of $\mathcal{Z}_{\varphi}(\mathcal{A}_{\geq n})$ 
is {\bf{continuous}}; (it is homeomorphic to the interval $[0,1]$). This is a new feature exhibited by the model considered 
here, which motivates one to generalize the collapse postulate, {\bf{Axiom CP}} of Sect.~\ref{ETH}, to systems 
featuring actual events, 
$$\big\{ \pi_{\xi}\,\vert \,\xi \in \mathfrak{X} \big\} \,\,\text{generating   }\,\, \mathcal{Z}_{\omega_{\,t}}\big(\mathcal{E}_{\geq t}\big),$$
(see Definition 6 of Sect.~\ref{ETH}) with a spectrum $\mathfrak{X}$ that can be a \textit{continuous} topological (compact Hausdorff) space. An extension of our theory to this situation will be pursued elsewhere.

We observe that, in the models considered in this section, an arbitrary projection $\pi^{(n)}$ in 
$\mathcal{Z}_{\varphi}(\mathcal{A}_{\geq n})$ has the form
$$\pi^{(n)} = P\vert_{\mathcal{H}_n} \otimes \pi_{\geq(n+1)},$$
where $P$ is a spectral projection of the density matrix $\Phi\vert_{\mathcal{H}_n}$ and $\pi_{\geq (n+1)} \in \mathcal{A}_{\geq(n+1)}$.\\ 
If $P= P_{k_1}+\dots + P_{k_J},$ for some \,$1< J\leq K,$ where the operators $P_{k_j}$ are spectral projections of 
$\Phi$, then $\pi^{(n)}$ can be decomposed into a non-trivial sum of projections,
$$\pi^{(n)}= \sum_{j=1}^{J} P_{k_j}\otimes \pi_{\geq(n+1)}\,.$$
Next, we recall the Collapse Postulate, {\bf{Axiom CP}}, in Sect.~\ref{ETH}.~In the context of the models discussed 
in this section it is natural to generalize this postulate as follows: 
If $\varphi$ is the initial state of the $R$-field then the state, $\varphi_n$, on the algebra $\mathcal{A}_{\geq n}$ visited, 
at time $n$, along some history of the system with initial condition $\varphi$ for the $R$-field has the form
\begin{equation}\label{phi-n}
\varphi_{n} = P_{k_n} \otimes \varphi^{(n+1)}\,, \quad \, \text{ for some   }\,\, k_n = 1,\dots K\,,
\end{equation}
where\,\, $\varphi^{(n+1)} $\,\,\ is a normal state on\,\, $\mathcal{A}_{\geq(n+1)}$. The frequency of choosing $k_n=k_{*}$,
for some $k_{*}\in \{1,\dots, K\}$, is given by
\begin{equation}\label{prob of branch}
\text{prob}(k_n=k_{*}) = p_{k_{*}}, \quad \text{ with  }\,\, \,p_{k_{*}} \,\, \text{ as in }\,\, \, \text{Eq.}\,\eqref{density matrix}\,.
\end{equation}
In the Schr\"odinger picture, the time evolution of the state of the atom coupled to the $R$-field predicted by the 
$ETH$-Approach (see Definition 6 and {\bf{Axiom CP}} of Subsect.~4.3) is then described by Eq.~\eqref{step-1-evol}, 
where the coefficients $g^{\ell m}$ are given by
\begin{equation}\label{stoch metric}
g^{\ell m}(j):= \big[\text{tr}(P_{k_j})\big]^{-1} \text{tr}\big(\, T^{(\ell)}\,P_{k_j}\, (T^{(m)})^{*} \big)\,.
\end{equation}
These coefficients are \textit{random variables} whose law is given by \eqref{prob of branch}.

It is clear that properties \eqref{phi-n}, \eqref{prob of branch} and \eqref{stoch metric} hold for sufficiently large times, $n$, for a subspace of initial states of the $R$-field dense in $\mathfrak{F}_{\varphi}$.
It would be interesting to analyze properties of the dynamics of the atom, with the \textit{randomness} in the time evolution of its states caused by the repeated collapse of the state of the $R$-field, as described in Eqs.~\eqref{phi-n} and 
\eqref{prob of branch}.

\section{Conclusions and outlook}\label{Conclusions}

\hspace{0.5cm}\textit{``The interpretation of quantum mechanics has been dealt with by many authors, and I do not want to discuss it here. I want to deal with more fundamental things.''} (Paul Adrien Maurice Dirac)\\

Our main goal in this paper has been to illustrate the rather abstract {\bf{$ETH$-Approach to Quantum Mechanics}} 
with a discussion of simple models, which are, however, sophisticated enough to exhibit some of the main subtleties 
and virtues of this approach. The models used in Sect.~\ref{Model} to illustrate the abstract ideas underlying the 
$ETH$-Approach have been inspired by implications of Huygens' Principle in quantum field theory and of the form 
it takes in the limit where the speed of light tends to $\infty$; see Sect.~\ref{HP}. Our main results are contained in
Sects.~\ref{ETH} and \ref{Model}.

To conclude this paper, we attempt to clarify what we consider to be the \textit{ontology} at the base of Quantum 
Mechanics, as suggested by the $ETH$-Approach. We then present some remarks and comments about the limits of 
the models in Sect.~\ref{Model} when time is a continuous parameter (i.e., the time step tends to 0). We conclude 
with some comments on different mechanisms that can give rise to the \textit{Principle of Diminishing Potentialities}. 
Along the way we draw the readers' attention to some important open problems.

\subsection{From `what may potentially be' to `what actually \textit{is}'}

\hspace{0.5cm}\textit{``The Garden of Forking Paths is a picture, incomplete yet not false, of the universe.''} (Jorge Luis Borges)

The summary of the $ETH$-Approach presented in Sect.~\ref{ETH} and the discussion of concrete models contained 
in Sect.~\ref{Model} provide a fairly clear idea of what might be considered to be the \textit{ontology} underlying Quantum 
Mechanics. In order to keep the following remarks as accessible as possible, we shall discuss this topic in the 
context of the models studied in the last section.

 Equation \eqref{event alg} of Sect.~\ref{Model} shows that, in the (idealized) models of physical systems studied 
 there, the event algebras $\mathcal{E}_{\geq n}, n\geq 0,$ are all unitarily equivalent to one ``universal'' algebra 
 $\mathcal{N}\equiv \mathcal{E} := \mathcal{A}_{\geq 0}\otimes B(\mathfrak{h}_S)$. (Recall that we only monitor 
 the evolution of the systems for times $t\geq t_{in}=0$.) The fact that 
 $\mathcal{E}_{\geq n} \simeq \mathcal{E}, \forall\, n\geq 0,$ enables us to define the 
 \textit{non-commutative spectrum}, $\mathfrak{Z}_S$, of the systems described by our models by setting
\begin{equation*}
\mathfrak{Z}_{S}:= \bigcup_{\omega} \Big(\,\omega, \mathcal{Z}_{\omega}(\mathcal{E})\Big)\,, 
\end{equation*}
where the union is a disjoint union ranging over all normal states $\omega$ on the algebra $\mathcal{E}$, and 
$\mathcal{Z}_{\omega}(\mathcal{E})$ is the center of the centralizer of the state $\omega$ restricted to the algebra 
$\mathcal{E}$; see Eq.~\eqref{NCspect} of Sect.~\ref{ETH}. The algebra $\mathcal{Z}_{\omega}(\mathcal{E})$ is abelian. 
Its lattice of projections provides a mathematical description of the actual event featured by the system when it 
occupies $\omega$.

Let $\gamma$ denote the $^{*}$-endomorphism of the algebra $\mathcal{E}$ corresponding to time translation 
of operators in the Heisenberg picture by a time step of length 1; i.e., 
$$\gamma(\mathcal{E}) = \mathcal{E}_{\geq 1} \subset \mathcal{E}\,.$$
\textit{Remark:} If time translations are unitarily implementable on the Hilbert space, $\mathcal{H}_S$, of state vectors
of the system $S$ then one has that 
$\gamma(X)= \Gamma^{-1} X\, \Gamma, \,\, \forall\,\, X\in \mathcal{E}$, where 
$\Gamma$ is the unitary propagator on $\mathcal{H}_S$ by a time step of length 1; see Sects.~\ref{ETH} and \ref{Model}.

Given the algebra $\mathcal{E}$ and the time-translation endomorphism $\gamma$ on $\mathcal{E}$, the space of normal states on $\mathcal{E}$ can be equipped with the structure of a \textit{groupoid}:

For a given pair, $(\omega, \omega')$, of normal states on $\mathcal{E}$, there is an \textit{arrow} from 
$\omega$ to $\omega'$, written as \,\,$\omega \rightarrow \omega'$, iff there exists a minimal orthogonal projection 
$\pi \in \mathcal{Z}_{\omega}(\mathcal{E})$ (i.e., $\pi$ cannot be decomposed into a sum of two or more non-zero projections 
belonging to $\mathcal{Z}_{\omega}(\mathcal{E})$) such that  
\begin{equation}\label{arrow}
\omega(\pi)>0,\quad \text{ and }\quad \omega'\big(X\big) = \big[\omega(\pi)\big]^{-1} \omega\big(\pi \,\gamma(X)\, \pi \big)\,, \qquad \forall\,\, X\in \mathcal{E}\,.
\end{equation}

\underline{Definition 9}: A \textit{history} of length $r$ is a connected path, 
$\underline{\omega}_{r}:= \big(\omega_0, \dots, \omega_{r}\big)\,,$ of states on $\mathcal{E}$, with the property that
\begin{equation}\label{history}
\hspace{3.5cm} \omega_j \rightarrow \omega_{j+1}\,, \, \, \,\forall \,\,j=0, \dots, r-1\,.\hspace{3.7cm} \square
\end{equation}
If $\underline{\omega}_{r}$ is a history of length $r$ then there exist minimal orthogonal projections $\pi_j \in \mathcal{Z}_{\omega_j}(\mathcal{E})$, with $\omega_{j}(\pi_j)>0, \text{ for }\, j=0, \dots, r-1$,  such that 
\begin{equation}\label{state recursion}
 \omega_{j+1}(X)= \big[\omega_{j}(\pi_j)\big]^{-1} \omega_{j}(\pi_j \, \gamma(X)\, \pi_j)\,, \quad\forall\,\, X\in \mathcal{E}\,,\,\forall\,\,j=0,\dots, r-1\,.
\end{equation}
Thus, a history $\underline{\omega}_r$ of length $r$ can also be parametrized by a pair 
$\big(\omega, \underline{\pi}_r \big)$, where $\omega=\omega_0$ is the inital state of the system, and the sequence 
of projections, $\underline{\pi}_r =\big(\pi_0, \dots, \pi_{r-1}\big)$ is such that Eq.~\eqref{state recursion} holds. The 
space of histories with initial condition $\omega\equiv \omega_0$ is denoted by $\mathfrak{H}_{\omega}$.
We define \textit{history operators}
\begin{equation}
H(\underline{\pi}_r):= \prod_{j=0}^{r-1}\gamma^{j}(\pi_j)\,, \quad \underline{\pi}_r =\big(\pi_0, \dots, \pi_{r-1}\big)\,,\quad r=1,2,3,\dots
\end{equation}
History operators can be used to equip $\mathfrak{H}_{\omega}$ with a probability measure, $\mathbb{P}_{\omega}$:
\begin{equation}\label{prob of history}
\mathbb{P}_{\omega}\big(\underline{\pi}_r\big):= \omega\big( H(\underline{\pi}_r)^{*}\cdot H(\underline{\pi}_r)\big)\,,\quad \omega =\omega_{0} \,,
\end{equation}
with $\pi_{j} \in \mathcal{Z}_{\omega_j}(\mathcal{E})$ and $\omega_j$ as in \eqref{state recursion}, for $j=0, \dots, r-1$. 
We have that
$$\sum_{\pi_{r-1} \in \mathcal{Z}_{\omega_{r-1}}(\mathcal{E})} \mathbb{P}_{\omega}\big(\underline{\pi}_r\big) = 
\mathbb{P}_{\omega}\big(\underline{\pi}_{r-1}\big)\,,$$
which follows readily from the definition of history operators, the fact that $\pi^{2}=\pi= \pi^{*}$, for an arbitrary orthogonal projection, and from the property that the projections $\pi_{r-1}\in \mathcal{Z}_{\omega_{r-1}}(\mathcal{E})$ form a partition of unity. \textit{Kolmogorov}'s extension lemma then tells us that $\mathbb{P}_{\omega}$ extends to a probability measure on the space $\mathfrak{H}_{\omega}$ of histories with initial condition $\omega$. 

Formula \eqref{prob of history} is reminiscent of the \textit{L\"uders-Schwinger-Wigner} formula \cite{LSW} for the probability of outcomes of repeated measurements; but it has a rather different status and interpretation.

\textit{The {\bf{ontology}} at the base of Quantum Mechanics lies in the {\bf{histories}} traversed by isolated physical systems.} More precisely, one might want to claim that what really ``exists'' is encoded into sequences
$$\Big\{ \Big(\omega_j, \mathcal{Z}_{\omega_j}(\mathcal{E}) \Big) \,\Big| \, \omega_j \rightarrow \omega_{j+1},\,\text{for }\, j=0, \dots, r-1\Big\}, \quad r=1,2,3,\dots\,,$$
with Eq.~\eqref{state recursion} providing the relation between $\omega_j, \pi_j$ and $\omega_{j+1}$.

We note that a state $\omega$ on the algebra $\mathcal{E}$ gives rise to an actual {\bf{E}}vent described by 
$\mathcal{Z}_{\omega}(\mathcal{E})$; the space of normal states on $\mathcal{E}$, viewed as a groupoid with arrow 
defined in \eqref{state recursion}, has a {\bf{T}}ree-like structure; and the states occupied by the system in the course of 
time form a {\bf{H}}istory, i.e., an element of the space $\mathfrak{H}_{\omega}$ of histories of the system starting in 
state $\omega$. This explains why the formulation of Quantum Mechanics explored in this paper is called 
{\bf{ETH}}-Approach. 

\underline{Problem}: Generalize the theory developed in Sect.~\ref{ETH} and exemplified by the models in Sect.~\ref{Model} to apply to physical systems with the following properties:
\begin{itemize}
\item{They have states, $\omega$, of physical interest that give rise to centers, $\mathcal{Z}_{\omega}(\mathcal{E})$, of centralizes with \textit{continuous spectrum}. (A preliminary version of such a generalization has been worked out and will appear elsewhere.)}
\item{Time is continuous, $t\in \mathbb{R}$.}
\item{They are described by \textit{relativistic local quantum theory}; (with `time' traded for `space-time'). A beginning of such a theory has been presented in \cite{Fr2}.}
\end{itemize}
We expect that the first problem stated here can be solved without major difficulties. Comments on the second and third problem follow in the next subsections.

\subsection{Models with continuous time}
\hspace{0.5cm}\textit{``Time does not pass, it continues.''} (Marty Rubin)

Recall the family of models discussed in Sect.~6. One should ask whether these models remain meaningful in the limit
where the time step tends to 0, i.e., for a continuous time parameter. 
To answer this question, we consider an $R$-field defined in terms of its  $\mathbb{M}_{N}(\mathbb{C})$-valued creation- and annihilation operators,
$a^{*}(t)$ and $a(t)$,  with 
\begin{align} \label{CCR}
a^{\#} = a\,& \text{ or }\, a^{*}, \qquad a^{\#}(t)= \big(a_{ij}^{\#}(t)\big)_{i,j=1,\dots, N}, \,\text{ and }\nonumber \\
\big[a_{ij}^{\#}(t), a_{k \ell}^{\#}(t')\big] =0,\qquad &\big[a_{ij}(t), a_{k\ell}^{*}(t')\big]= \delta_{ik}\,\delta_{j\ell}\cdot \delta(t-t')\,, \quad \forall\,\, i,j,k, \ell, \quad \forall\,\,t \,, t' \text{ in }\mathbb{R}\,. 
\end{align}
Let $\mathfrak{F}$ denote the Fock space corresponding to these creation- and annihilation operators; the creation- and annihilation operators, $a_{ij}^{\#}(\cdot),$ are operator-valued distributions on $\mathfrak{F}$. Fock space contains a vector $\vert 0 \rangle$, called \textit{vacuum vector}, with the property
$$a_{ij}(t)\vert 0 \rangle = 0, \qquad \forall \,\, i,j = 1,\dots, N, \,\, \forall\,\, t\in \mathbb{R}\,.$$
 Applying arbitrary polynomials in creation operators, smeared out with $\mathbb{M}_{N}(\mathbb{C})$-valued 
test functions on the time axis $\mathbb{R}$, to the vacuum vector $\vert 0\rangle$ generates a dense set of vectors 
in $\mathfrak{F}_S$. Fourier transformation in the variable $t$ yields creation- and annihilation operators, $\hat{a}^{*}(\nu)$ 
and $\hat{a}(\nu)$, related to $a^{*}(t)$ and $a(t)$ by
$$a_{ij}^{\#}(t)= \frac{1}{\sqrt{2\pi}} \int_{\mathbb{R}} d\nu\,e^{\pm i t\cdot \nu}\, \hat{a}_{ij}^{\#}(\nu),$$
and satisfying the commutation relations \eqref{CCR}, with time $t$ replaced by frequency $\nu$.
Time translations on $\mathfrak{F}$ are generated by an operator
\begin{equation}\label{time transl}
H_R := \frac{1}{2\pi} \sum_{i.j=1}^{N}\int_{\mathbb{R}} d\nu\, \hat{a}_{ij}^{*}(\nu)\, \nu \,\hat{a}_{ij}(\nu)\,.
\end{equation}
The Hamilton operator $H_R$ is self-adjoint on a natural dense subspace of $\mathfrak{F}$ and generates 
unitary time translations,
$$\Gamma_t = e^{-it\,H_R}$$
on $\mathfrak{F}$. We observe that the spectrum of $H_R$ covers the entire real axis,\footnote{It is a general theorem that the existence of \textit{time-translation invariant product states} implies that the spectrum of the Hamiltonian is unbounded from above and from below} that the vacuum vector $\vert0\rangle$ is invariant under time-translations, i.e., 
$e^{itH_R}\vert 0 \rangle = \vert 0 \rangle,\, \forall \, t$, and that it is a product state. This last property follows from the form of the two-point function,
$$\langle0\vert a_{ij}(t)\, a^{*}_{k\ell}(t')\vert0\rangle = \delta_{ik}\delta_{j\ell} \delta(t-t')\,,$$
and Wick's theorem (see, e.g., \cite{Derezinski}).

Next, we add the ``atom'' to the play and introduce interactions between the atom and the $R$-field. We continue to simplify matters by assuming that the atomic Hilbert space is finite-dimensional, $\mathfrak{h}_S\simeq \mathbb{C}^{M},$ for some $M< \infty$. The Hilbert space for the atom coupled to the $R$-field is given by $\mathcal{H}_S = \mathfrak{F}\otimes \mathfrak{h}_S$. Before it is coupled to the $R$-field the propagator of the atom is generated by a hermitian matrix, $H_A$, on $\mathfrak{h}_S$, which we assume to be bounded. The interaction between the atom and the $R$-field is specified by a self-adjoint operator, $W=W^{*}$, 
on $\mathcal{H}_S$. The total Hamiltonian of the system $S$ is then given by
\begin{equation}\label{Ham}
H:= H_R\otimes \mathbf{1} + \mathbf{1}\otimes H_A + W
\end{equation}
The limit of continuous time of the model studied in Subsect.~6.4 corresponds to the choice
$$W= \sum_{m=1}^{L} v_m\otimes Q_m\,,$$
where $\big\{Q_m\big\}_{m=1}^{L}$ is a partition of unity on $\mathfrak{h}_S$ by orthogonal projections, as in Subsect~6.4, and the operators $v_m$ are self-adjoint bounded operators on $\mathfrak{F}$, (with $e^{-iv_m} = T^{(m)}$). 

Models of this sort have been studied in the literature; see, e.g., \cite{H-P, FGH} and refs.~given there.\footnote{They have also come up in connection with the problem of \textit{time} in QM and quantum systems describing \textit{clocks}; see \cite{clocks}. For us, the literature on this topic has been rather confusing.} Before the Collapse Postulate, {\bf{Axiom CP}} of Sect.~\ref{ETH}, is imposed the effective time evolution of the atomic degrees of freedom is given by a Lindbladian evolution \cite{GKS, Lindblad}. In \cite{Bassi}, the authors derive a non-linear stochastic Schr\"odinger equation for the state vector -- the state of the atom in our model -- from Lindbladian evolution of the density matrix when the Collapse Postulate is imposed. Their results can be applied to the model introduced above.

One might argue that one should attempt to derive continuous-time limits of the more natural (semi-relativistic) models studied in Sect.~\ref{HP}, which could be expected to have Hamiltonians that are bounded from below. However, this project is obstructed by our inability to construct models of \textit{local relativistic quantum theory}, in particular Quantum Electrodynamics, without ultraviolet cutoffs. Thus, in the realm of (semi-)relativistic models of atoms coupled to the quantized electromagnetic field satisfying the Principle of Diminishing Potentialities, we are stuck with models that have a discrete time, as discussed in Sect.~\ref{HP}. (See, however, \cite{Fr2} for an ``axiomatic'' analysis of the $ETH$-Approach to local relativistic quantum theory.)

\subsection{Are there alternatives to Huygens' Principle in deriving the Principle of Diminishing Potentialities?}

\hspace{0.5cm}\textit{``One finds in this subject a kind of demonstration which does not carry with it so high a degree of certainty as that employed in geometry,...''} (Christiaan Huygens)

In this last subsection, we pose the problem to identify physical mechanisms that give rise to the 
\textit{Principle of Diminshing Potentialities} $(PDP)$ (see Eq.~\eqref{PDP}, Sect.~\ref{ETH}). 
We have seen in Sect.~\ref{HP} that $(PDP)$ is implied by \textit{Huygens' Principle} \cite{Buchholz} in local relativistic 
quantum theories involving \textit{massless modes} and by the form this principle takes in quantum theories obtained 
in the limit of the speed of light tending to $\infty$. This suggests to study the question on what space-times 
Huygens' Principle is known to be valid.
\begin{itemize}
\item{Huygens' Principle is known to hold in theories on Minkowski space-times, $\mathbb{M}^{d}$, of \textit{even} dimension, 
i.e., for even $d$; and it is known to fail in theories on odd-dimensional Minkowski space-times. }
\item{Huygens' Principle holds on even-dimensional space-times diffeomorphic to (a positive-time halfspace in) 
$\mathbb{M}^{d}$ with a metric that differs from the standard Lorentzian metric on $\mathbb{M}^{d}$ only by a 
\textit{conformal factor}. An example is the spatially flat Friedman-Lema\^itre universe.\\
It would be of interest to compile a list of space-times on which Huygens' Principle holds true.}
\item{We expect that $(PDP)$ holds on certain even-dimensional space-times with black holes. But we have not studied 
this issue in any detail, yet.}
\item{There are even-dimensional space-time manifolds with non-vanishing curvature on which Huygens' Principle 
\textit{fails}. However, this may not invalidate $(PDP)$, as remarked next.}
\item{Huygens' Principle and $(PDP)$ could hold if it turned out that ``visible'' space-time is a submanifold of positive co-dimension of a space-time manifold with \textit{extra dimensions}, and only certain massless modes could and would penetrate into the bulk of the higher-dimensional space-time manifold} (\textit{even} if, on the submanifold corresponding to the ``visible'' space-time, Huygens' Principle may fail).
\end{itemize}

The Principle of Diminishing Potentialities constrains the ``relative positions'' (inclusions) of algebras generated by 
\textit{potential events/potentialities} localized in the future of different causally ordered points in space-time; 
see Sect.~\ref{HP} and \cite{Fr2}. If gravity is neglected it is clear what is meant by the future of a space-time 
point $P$: It is the future light cone, $V^{+}_P$, erected over $P$, and the potentialities localized in the future of 
$P$ are certain operators localized in $V^{+}_P$ that generate an algebra denoted by $\mathcal{E}_{\geq P}$. 
The bundle of future light cones over space-time is determined by the conformal structure of space-time, and 
Huygens' Principle is tied to properties of the propagation of (massless) waves on space-time.

However, if gravitational effects are taken into account the structure of future light-cones in space-time and the metric 
in the vicinity of future light-cones are not determined a priori, because quantum theory does never determine with certainty 
\textit{what events/actualities} will happen. Since events couple to gravity, the metric structure of the ``future'' is 
not determined a priori.  For these reasons, the Principle of Diminishing Potentialities should really be formulated in 
a way that does \textit{not} depend on knowledge of the conformal structure in the vicinity of future light-cones in space-time. 
One should look for a more abstract, \textit{background-independent} formulation of $(PDP)$, one that incorporates 
gravitational effects and, in particular, the role gravity plays in making $(PDP)$ possible.

We recall that one expects that, in a given local relativistic quantum theory, all event algebras,
$\mathcal{E}_{\geq P}$, associated with the future above an arbitrary space-time point $P$ are isomorphic to a 
universal algebra $\mathcal{N}$. The Principle of Diminishing Potentialities can then be seen as a consequence of 
the existence of one-parameter semi-groups, $\big\{\gamma_t\big\}_{t\in [0, t_{*})}\,, 0< t_{*} \leq \infty$, of 
$^{*}$-endomorphisms of $\mathcal{N}$ with the property that 
\begin{equation}\label{abstract endo}
\gamma_{t}\big(\mathcal{N}\big) \subsetneqq \mathcal{N}, \qquad \forall\,\,t>0\,.
\end{equation}
The problem is to come up with a general characterization of algebras that can play the role of $\mathcal{N}$ 
and of one-parameter semi-groups $\big\{\gamma_t\big\}_{t\in [0, t_{*})}$ on such algebras satisfying
\eqref{abstract endo}. This problem appears to be a very difficult one.

Returning to quantum theories on Minkowski space, with gravity neglected, our analysis leads to the following somewhat 
tantalizing general {\bf{conjecture}}:
If we consider a quantum theory for a system $S$ in which $(PDP)$ holds and with a Hamiltonian, 
$H$, generating Heisenberg-picture time translations of operators representing physical quantities of $S$ 
that satisfies the spectrum condition, i.e., $H\geq 0$, then this theory must necessarily be a  \textit{local relativistic} 
quantum theory on an even-dimensional Minkowski space. In other words, a quantum theory describing \textit{events} 
and \textit{measurements}, which does not have states of arbitrarily negative energy, must be a local relativistic theory 
on an even-dimensional space-time.\\

To conclude this discussion, one might say that the Principle of Diminishing Potentialities $(PDP)$ is really the appropriate 
general formulation of Huygens' Principle in local quantum theory, and that there may not be any viable alternatives to $(PDP)$ if we 
want quantum theory to describe events (including measurements and observations). Thus, a clarification of the status of the Principle of Diminishing Potentialities may be viewed as a fundamental problem of Quantum Physics.

\end{document}